\documentclass{aa}  

\usepackage{savesym}
\usepackage{amsmath}
\savesymbol{iint}
\usepackage{txfonts}
\restoresymbol{TXF}{iint}
\usepackage{graphics,graphicx}
\usepackage{amssymb}
\usepackage{color}
\usepackage[draft]{hyperref}
\usepackage{breakurl}
\usepackage{array}
\usepackage{rotating}
\usepackage{xcolor}
\usepackage{natbib}
\usepackage{float}
\usepackage{flafter}
\usepackage{booktabs}
\usepackage{afterpage}
\usepackage{lipsum}
\usepackage{longtable}
\usepackage{lscape}
\bibpunct{(}{)}{;}{a}{}{,}

\newcommand{\obj}{30~Dor~C}

\newcommand{\chandra}{{\it Chandra}}
\newcommand{\xmm}{{\it XMM-Newton}}

\newcommand{\hess}{H.E.S.S.}

\newcommand{\ha}{H$\alpha$}

\newcommand{\nii}{[N$_{\rm{II}}$]}

\newcommand{\bfield}{{\it B}-field}

\begin{document}

   \title{Magnetic field estimates from the X-ray synchrotron emitting rims of the 30 Dor C superbubble and the implications for the nature of 30 Dor C’s TeV emission}

   \author{ Patrick~J.~Kavanagh\inst{1} 
            \and 
            Jacco Vink\inst{2,3,4} 
            \and 
            Manami~Sasaki\inst{5}
            \and 
            You-Hua Chu\inst{6} 
			\and 
            Miroslav~D.~Filipovi\'c\inst{7}
            \and
            Stefan Ohm\inst{8} 
            \\
            Frank~Haberl\inst{9}
            \and 
            Perica~Manojlovic\inst{7}
            \and
            Pierre Maggi\inst{10} 
          }


\institute{
School of Cosmic Physics, Dublin Institute for Advanced Studies, 31 Fitzwillam Place, Dublin 2, Ireland, \email{pkavanagh@cp.dias.ie}
\and
Anton  Pannekoek  Institute  for  Astronomy,  University  of  Amsterdam, Science Park 904, 1098 XH Amsterdam, The Netherlands
\and
GRAPPA, University of Amsterdam, Science Park 904, 1098 XH Amsterdam, The Netherlands
\and
SRON, Netherlands Institute for Space Research, Utrecht, The Netherlands
\and 
Remeis Observatory and ECAP, Universit\"{a}t Erlangen-N\"{u}rnberg, Sternwartstr. 7, D-96049 Bamberg, Germany
\and
Institute of Astronomy and Astrophysics, Academia Sinica (ASIAA), Taipei 10617, Taiwan
\and
Western Sydney University, Locked Bag 1791, Penrith, NSW 2751, Australia
\and
DESY, D-15738 Zeuthen, Germany
\and
Max-Planck-Institut f\"{u}r extraterrestrische Physik, Giessenbachstra\ss e, D-85748 Garching, Germany
\and
Laboratoire AIM, IRFU/Service d'Astrophysique - CEA/DRF - CNRS - Universit\'{e} Paris Diderot, Bat. 709, CEA-Saclay, 91191 Gif-sur-Yvette Cedex, France
}

\titlerunning{Magnetic field estimates in 30 Dor C}

\date{Received --; accepted --}

\abstract{The \obj\ superbubble is unique for its synchrotron X-ray shell, as well as being the first superbubble to be detected in TeV $\gamma$-rays, though the dominant TeV emission mechanism, i.e., leptonic or hadronic, is still unclear.}{We aim to use new \chandra\ observations of \obj\ to resolve the synchrotron shell in unprecedented detail and to estimate the magnetic ($B$) field in the postshock region, a key discriminator between TeV $\gamma$-ray emission mechanisms.}{We extracted radial profiles in the 1.5--8~keV range from various sectors around the synchrotron shell and fitted these with a projected and point spread function convolved postshock volumetric emissivity model to determine the filament widths. We then calculated the postshock magnetic field strength from these widths.}{We found that most of the sectors were well fitted with our postshock model and the determined \bfield\ values were low, all with best fits $\lesssim 20~\mu$G. Upper limits on the confidence intervals of three sectors reached $\gtrsim 30~\mu$G though these were poorly constrained. The generally low \bfield\ values suggests a leptonic-dominated origin for the TeV $\gamma$-rays. Our postshock model did not provide adequate fits to two sectors. We found that one sector simply did not provide a clean enough radial profile, while the other could be fitted with a modified postshock model where the projected profile falls off abruptly below $\sim0.8$ times the shell radius, yielding a postshock \bfield\ of 4.8~(3.7--11.8)~$\mu$G which is again consistent with the leptonic TeV $\gamma$-ray mechanism. Alternatively, the observed profiles in these sectors could result from synchrotron enhancements around a shock-cloud interaction as suggested in previous works.}{The average postshock \bfield\ determined around the X-ray synchrotron shell of \obj\ suggests the leptonic scenario as the dominant emission mechanism for the TeV $\gamma$-rays.}

\maketitle

\section{Introduction}
\label{intro}
Superbubbles (SBs) are large, $100-1000$~pc diameter
shells of swept-up interstellar medium (ISM) which are carved by the mechanical output of massive star clusters, i.e., via stellar winds and supernovae (SNe). The interior of these shells is filled with a hot ($10^{6}$~K), shock-heated gas
\citep[e.g.,][]{MacLow1988} while the swept-up shell of material is revealed by photo-ionisation of the shell by the photon field of the driving massive stellar population. \object{30~Dor~C} in the \object{Large Magellanic Cloud} (\object{LMC}) is unique among SBs as it exhibits a
bright non-thermal X-ray shell \citep{Dennerl2001}. The hard
shell was found to be synchrotron in origin by \citet[][henceforth BU04]{Bamba2004}, \citet[][henceforth SW04]{Smith2004}, \citet{Yam2009},
and \citet[][henceforth KS15]{Kavanagh2015},
indicating the presence of very high-energy (VHE) electrons. It has been suggested by BU04,
\citet{Yam2009}, and \citet[][hereafter HC15]{hess2015} that the synchrotron X-ray emission is due to a rapidly expanding
SNR in \obj. 

\citet[][hereafter SY17]{Sano2017} identified molecular material associated with \obj\ using Mopra observations of the $^{12}$CO line, with the brightest CO clouds distributed along the western shell. Comparing the radial profiles of the synchrotron X-rays and CO revealed an apparent X-ray excess around the CO peaks on a 10~pc scale, and CO peaks offset from X-ray peaks on a 1~pc scale. SY17 suggested that this correlation between synchrotron X-rays and molecular clouds is an indication of a shock-cloud interaction \citep{Inoue2009,Inoue2012}, similar to some Galactic SNRs \citep[e.g.,][]{Sano2010,Sano2013}, and that the observed non-thermal X-ray emission is the result of VHE electrons losing energy in the high, amplified $\sim$mG magnetic fields of the turbulent shock-cloud interaction region.

The detection of synchrotron X-rays in \obj\ reveals the presence of VHE electrons up to at least $10^{13}$~eV, and indicates that particle acceleration is ongoing in the SB. The recent detection of TeV $\gamma$-rays from \obj\ by the High Energy Stereoscopic System (\hess) has shown that accelerated particles in SBs give rise to TeV emission (HC15). This detection, the first at such energies, identified SBs as a new and
important source class in TeV astronomy. However, the dominant production mechanism of the TeV $\gamma$-rays (i.e., hadronic or leptonic) remains unclear. 

\begin{figure*}[!ht]
\begin{center}
\resizebox{\hsize}{!}{\includegraphics[trim= 0cm 0cm 0cm 0cm, clip=true, angle=0]{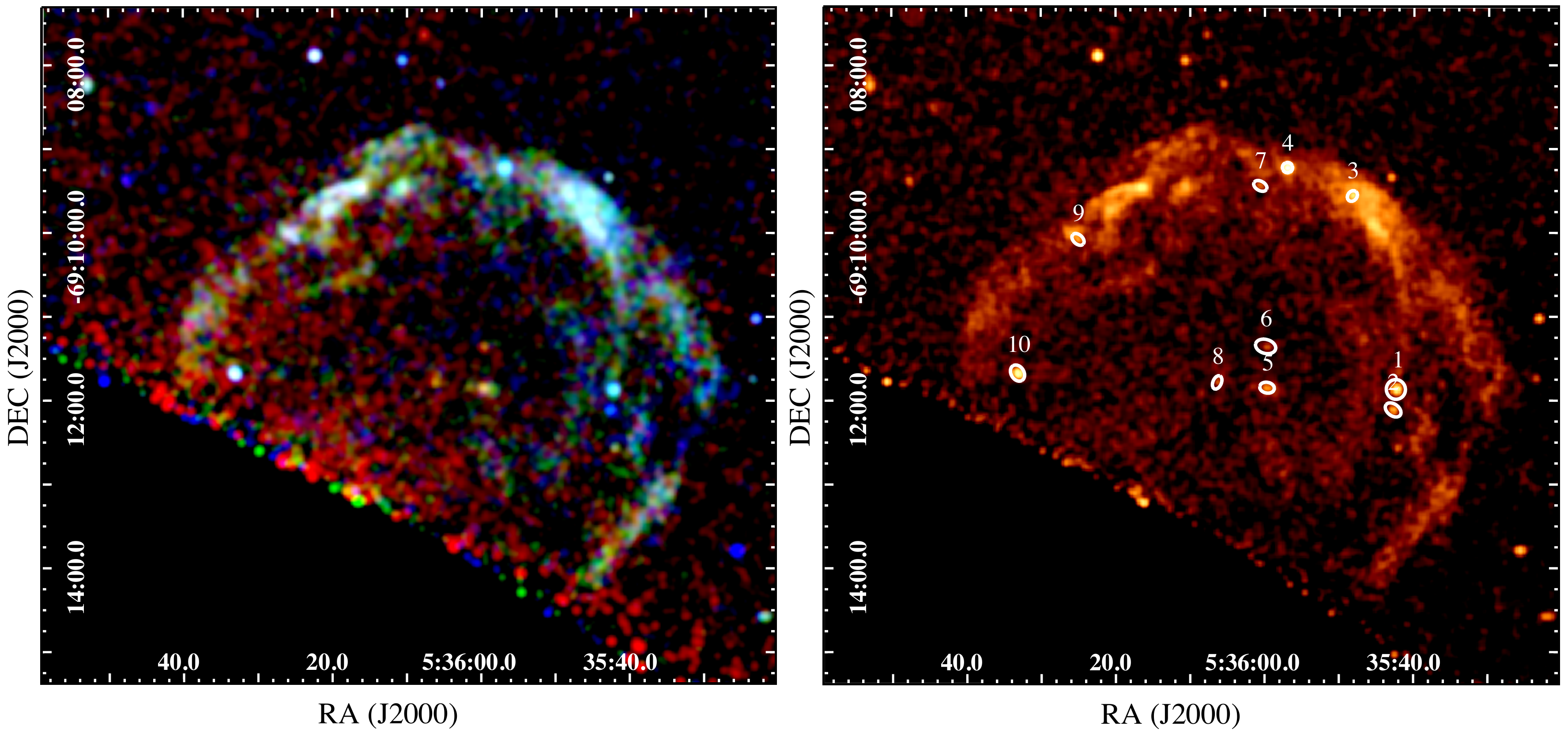}}
\caption{{\it Left:} False-colour \chandra\ image of \obj\ with RGB = 0.3--1~keV, 1--2~keV, 2--8~keV. {\it Right:} 0.5--8~keV image of \obj\ with detected sources overlaid. Source properties are given in Table~\ref{table:chandra_sources}. The images have been smoothed using a $3\sigma$ Gaussian kernel.}
\label{figure:chandra_images}
\end{center}
\end{figure*}

The \hess\ source (HESS~J0535-691) located in \obj\ 
has a measured J2000 position of RA~=~05:35:$(55\pm5)$,
Dec~=~--69:11:$(10\pm20)$ and a 1--10~TeV $\gamma$-ray luminosity of
$(0.9\pm0.2)\times10^{35}$~erg~s$^{-1}$ (HC15). The best-fit position is
located between the six identified sub-clusters \citep{Lortet1984} of the LH~90 OB association,
shifted towards the synchrotron shell. TeV $\gamma$-ray emission can result
from either the production of neutral pions via the collision of hadronic
cosmic rays with ambient material (hadronic scenario) or from inverse Compton
(IC) scattering of low-energy photons to $\gamma$-ray energies by VHE
electrons\footnote{The use of `electrons' here refers to both electrons and
positrons.}. HC15 could not definitively determine the
dominant mechanism responsible for the $\gamma$-ray emission, with both
hadronic and leptonic scenarios possible under certain conditions. Similarly, SY17 applied both hadronic and leptonic models in light of the possible shock-cloud interaction regions in \obj\ and could not rule out a hadronic or leptonic scenario.

A key discriminator between the hadronic and leptonic models is the average strength of the magnetic field ($B$) downstream of the shock. For a purely leptonic scenario, both HC15 and SY17 models require an average $B$-field $\sim15$~$\mu$G, whereas a higher, amplified $B$-field is required for a hadronic-dominated scenario to account for the observed synchrotron X-ray emission. The latter depends on the assumed energy in electrons for a fixed set of input parameters for the protons (i.e. the e/p ratio) and, as we show in Sect.~\ref{sect:hadronic_bfield}, $\gtrsim50~\mu$G suggests the leptonic contribution is insignificant, and $\sim20~\mu$G suggests a comparable contribution. Therefore, the \bfield\ in \obj\ is a crucial piece of the puzzle regarding the relative contributions to the TeV $\gamma$-ray emission from \obj. 

The strength of the downstream \bfield\ can be estimated from the widths of the synchrotron X-ray filaments. These widths are determined by synchrotron energy losses in combination with transport (diffusion and advection) of the electrons downstream of the shock: while they are being advected away from the shock, electrons may, after some time, have lost so much energy that they are no longer energetic enough to emit X-rays \citep[e.g.,][and references therein]{Reynolds1981,Vink2012}. The time scale for energy losses ($\tau_\mathrm{syn}$) is inversely proportional to the $B$-field strength, which translates into an advection length scale of $l_\mathrm{adv}=v_2 \tau_\mathrm{syn}$,
with $v_2=v_\mathrm{s}/4$ the downstream advection velocity.
Near the maximum electron energy, where synchrotron losses are balanced by acceleration gains, the 
advection length scale becomes comparable to the diffusion length scale, $l_\mathrm{diff}=D_2/v_2$, with
$D_2$ the downstream diffusion coefficient, which is also inversely proportional to $B$. In that case, the
width of the synchrotron filaments becomes $\sqrt{D_2 \tau_\mathrm{syn}}$, and since $D\propto E$ and 
$\tau_\mathrm{syn}\propto E^{-1}$, the width will be a direct probe of the magnetic field strength and is 
independent of the electron energy and the advection velocity \citep[e.g.][]{Voelk2005,Vink2006,Helder2012,Rettig2012,Ressler2014}.

In this paper, we present new Chandra X-ray Observatory (\chandra) observations of the synchrotron shell of \obj\ which provide the sharpest view of the X-ray shell to date. This allowed us to investigate the shell morphology in unprecedented detail and estimate the \bfield\ from the synchrotron filaments. In addition, we present high-resolution optical images obtained in the Magellanic Cloud Emission Line Survey 2 (MCELS2) and 6~cm radio continuum data from the Australia Telescope Compact Array (ATCA) to provide the best view of the optical and radio shells to date, allowing us to perform a multi-wavelength morphological study to determine whether the X-ray, optical, and radio shells are correlated.

We present our observations and data reduction in Sect.~2 before presenting our analysis of these data in Sect.~3. We discuss the results of our analysis in Sect.~4 and offer our conclusions in Sect. 5.

\section{Observations and data reduction}
\subsection{Chandra}
\label{section:chandra-obs}
\obj\ was observed by \chandra\ \citep{Weiss1996} on 2017 May~3 (Obs. ID 17904, PI P.~J. Kavanagh) and 2017 May~12 (Obs. ID 19925, PI P.~J. Kavanagh) with the Advanced CCD Imaging Spectrometer S-array \cite[ACIS-S,][]{Garmire2003} as the primary instrument. In each observation, the brightest region of the synchrotron X-ray shell was placed at the aimpoint of the ACIS-S array on the back-illuminated S3 CCD. The front-illuminated S2 and S4 CCDs were also switched on in each observation. The telescope roll-angle for the observations differed slightly at $\sim205$~degrees and $\sim213$~degrees for Obs.~IDs 17904 and 19925, respectively.

We reduced and analysed the \chandra\ observations using the CIAO~v4.9\footnote{See \burl{http://cxc.harvard.edu/ciao/}} \citep{Fru2006} software package with CALDB~v4.5.9\footnote{See \burl{http://cxc.harvard.edu/caldb/}}. Each dataset was reduced using the contributed script \texttt{chandra\_repro}, resulting in filtered exposure times of 40.5~ks and 40.6~ks for Obs. IDs 17904 and 19925, respectively. We reprojected the level 2 event files from each observation to a common tangent point using the CIAO task \texttt{reproject\_obs} and merged the resulting event files. Fluxed images were produced in the 0.3--1~keV, 1--2~keV, 2--8~keV, 0.5--8~keV, and 1.5--8~keV energy ranges using the CIAO \texttt{fluximage} task. These images were used to create the three-colour composite image which is shown in Fig.~\ref{figure:chandra_images}-left.

Source detection was performed on the merged event file in the 0.5--8 keV range using the \texttt{wavdetect} task. This works by correlating the input image with a series of Mexican hat wavelets. For point sources, the optimum wavelet size or scale is comparable to the size of the point spread function (PSF). To compute the PSF map for the merged event file we generated PSF files for each of the observations individually using the CIAO task \texttt{mkpsfmap}. These were then combined and weighted according to the corresponding exposure maps. The output source list was examined to identify spurious sources associated with the extended emission in the FOV. After removing these, we were left with 10 sources located in \obj\ which are shown in Table \ref{table:chandra_sources} and overlaid on the 0.5--8~keV image in Fig. \ref{figure:chandra_images}-right.

\begin{table*}
\caption{Detected \chandra\ sources located in \obj.}
\begin{center}
\label{table:chandra_sources}
{\renewcommand{\arraystretch}{1.3}
\begin{tabular}{llllllll}
\hline
\hline
Source & RA & Dec & Cts. & Cts. err. & Rate & Rate err. & Signif. \\
 & (J2000) & (J2000) &  &  & ($10^{-6}$~s$^{-1}$) & ($10^{-6}$~s$^{-1}$) & ($\sigma$) \\
\hline
1 & 05:35:42.4 & -69:11:52.3 & 142.4 & 13.5 & 4.6 & 0.4 & 19.2 \\
2 & 05:35:42.8 & -69:12:06.9 & 46.8 & 8.1 & 1.7 & 0.3 & 8.7 \\
3 & 05:35:48.3 & -69:09:33.9 & 157.8 & 13.9 & 5.1 & 0.5 & 22.6 \\
4 & 05:35:57.0 & -69:09:13.6 & 237.8 & 16.2 & 7.7 & 0.5 & 39.0 \\
5 & 05:35:59.7 & -69:11:51.1 & 58.8 & 9.5 & 1.9 & 0.3 & 8.8 \\
6 & 05:35:59.9 & -69:11:21.6 & 23.4 & 6.6 & 0.8 & 0.2 & 4.2 \\
7 & 05:36:00.6 & -69:09:26.7 & 24.8 & 6.9 & 0.8 & 0.2 & 4.3 \\
8 & 05:36:06.4 & -69:11:47.3 & 18.8 & 5.6 & 0.6 & 0.2 & 4.0 \\
9 & 05:36:25.0 & -69:10:05.0 & 41.6 & 7.6 & 1.4 & 0.3 & 8.3 \\
10 & 05:36:33.2 & -69:11:40.6 & 154.5 & 15.4 & 5.6 & 0.6 & 15.3 \\

\hline
\end{tabular}}
\end{center}
\end{table*}%

\subsection{Optical}
\subsubsection{MCELS2}
\begin{figure*}
\begin{center}
\resizebox{\hsize}{!}{\includegraphics[trim= 0cm 0cm 0cm 0cm, clip=true, angle=0]{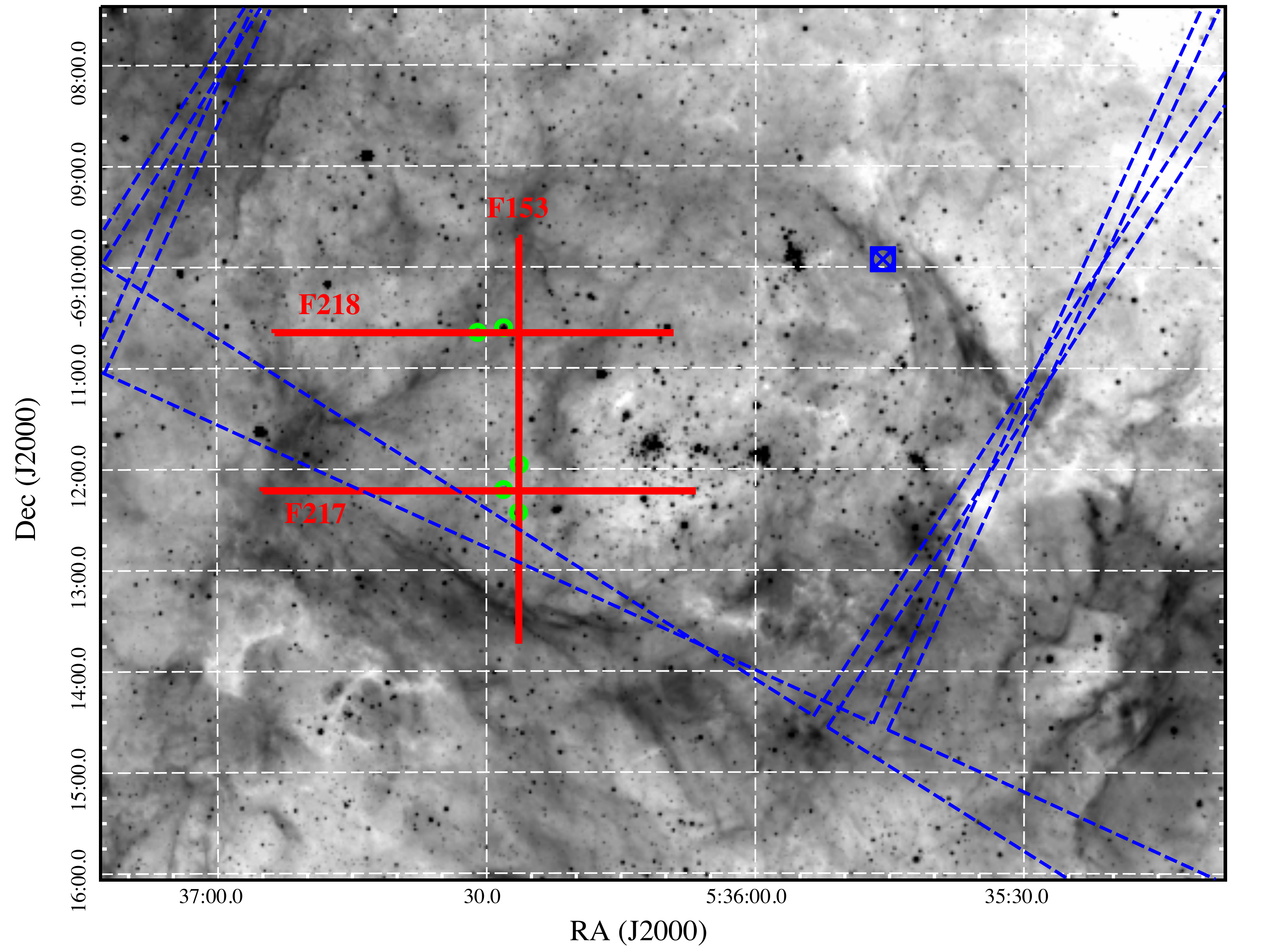}}
\caption{MCELS2 \ha\ image of \obj. The positions of optical spectroscopic slits are indicated by the red lines. The contaminating point sources are marked by the green circles. The blue circle-cross point marks the aim-point of the \chandra\ observations with the blue dashed lines delineating the ACIS-S coverage.}
\label{figure:slit_positions}
\end{center}
\end{figure*}

We made use of \ha\ images from MCELS2. The MCELS2 was performed with the Cerro Tololo Inter-American Observatory (CTIO) Blanco  4~m  telescope which used the MOSAIC II camera and covered the entire LMC. The MOSAIC II camera consists of eight SITe 4096$\times$2048 CCDs with a pixel size of $0.27\arcsec \times 0.27\arcsec$ and a combined field-of-view of $36\arcmin\times36\arcmin$. The SuperMACHO pipeline software was used for bias subtraction, flat-fielding, and distortion correction. The MCELS2 \ha\ image of \obj\ is shown in Fig.~\ref{figure:slit_positions}.

\subsubsection{Spectroscopy}
\label{opt-spec-intro}
\begin{figure*}[!ht]
\begin{center}
\resizebox{\hsize}{!}{\includegraphics[trim= 0cm 0cm 0cm 0cm, clip=true, angle=0]{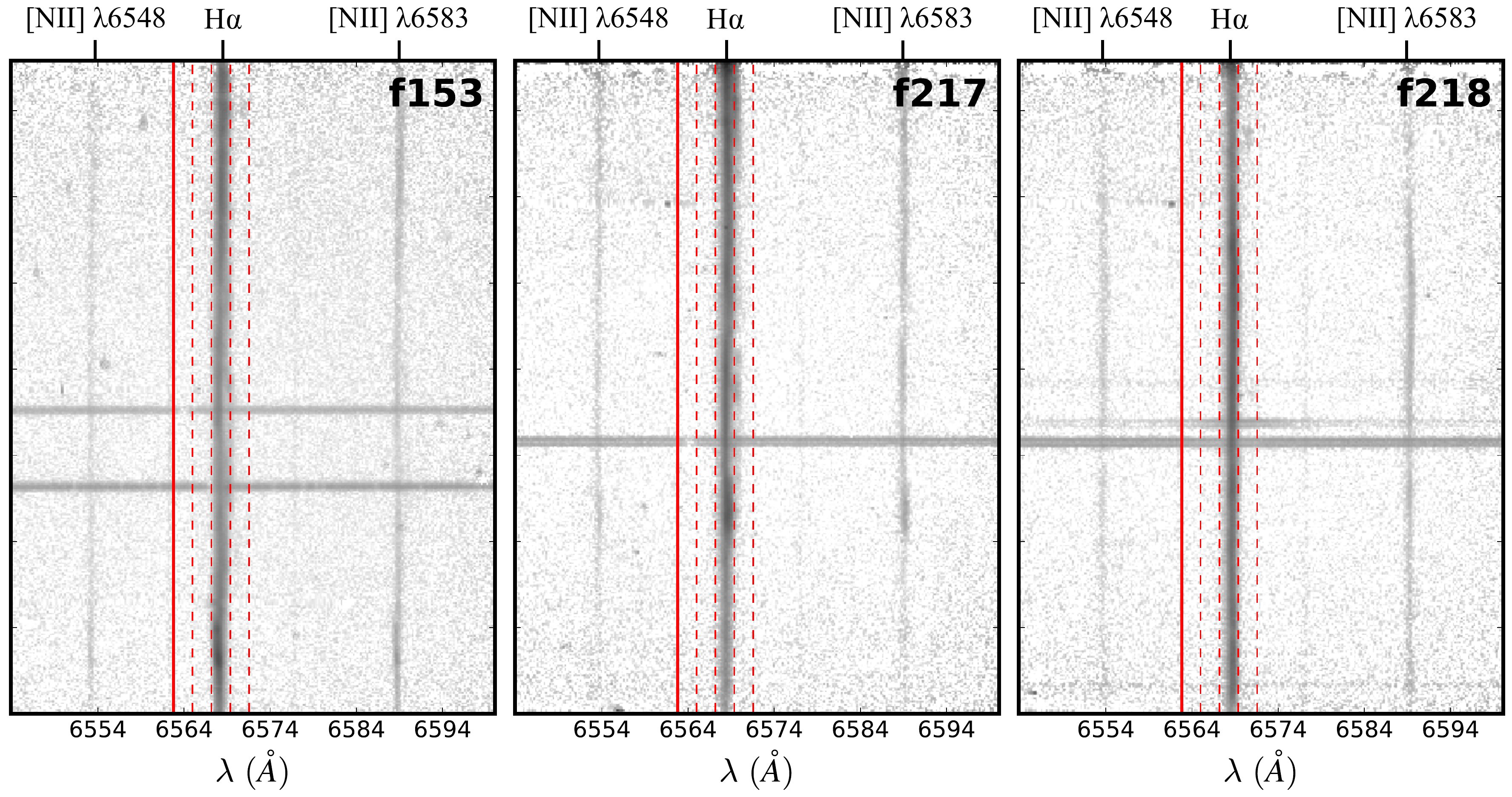}}
\caption{Spectro-images for each of the slit positions in Fig.~\ref{figure:slit_positions} from \citet{Chu1997}. Emission lines from \ha\ and \nii\ are indicated. The solid red line indicates the rest wavelength of \ha, while the dashed red lines indicate redshifted positions with recessional velocities from 100--400~km~s$^{-1}$ in steps of 100~km~s$^{-1}$. Note that the horizontal lines in each image are the continua of stars in the slits, marked by the green circles in Fig.~\ref{figure:slit_positions}.}
\label{figure:spectro-images}
\end{center}
\end{figure*}

Long-slit spectroscopy of the \obj\ shell has been performed in the past \citep{Chu1988, Chu1997}. To aid in our analysis and discussion, we made use of the spectroscopic data of \citet{Chu1997}. These data were obtained using the spectrograph on the CTIO Blanco 4~m telescope. Two slits were aligned in the east-west direction with one aligned in the north-south direction, as showing in Fig.~\ref{figure:slit_positions}. The data were reduced using the standard IRAF tasks to produce the spectro-images shown in Fig.~\ref{figure:spectro-images}.  The \ha\ line is clearly visible in the spectro-images at $\approx 6569$~\AA. Emission lines from \nii~$\lambda6548$ and \nii~$\lambda6583$ are also observed and are indicated in Fig.~\ref{figure:spectro-images}. In addition, continua from stars located in the slits are evident as the horizontal lines in the spectro-images. The contaminating stars are indicated in Fig.~\ref{figure:slit_positions}.

\subsection{Radio continuum}
We searched the Australia Telescope Online Archive\footnote{\burl{https://atoa.atnf.csiro.au/}} database for high resolution, and high dynamic range observations of this region. \obj\ is about $5\arcmin$ away from SN1987A which is one of the most frequently observed ATCA sources.  We used ATCA project C015 and we reduced the most recent (publicly available) ATCA CABB observations at 6~cm (5.5~GHz) which spans dates between 2010 and 2014. The observations totaled 77.68~hours time on source over 10 separate days. These observations are centered at SN1987A \citep{Ng2013} with a number of different arrays including 6A, 6B, and 1.5A. The data reduction was done using the MIRIAD software package \citep{Sault1995} and the final image used a Briggs robust weighting of 0.5 toward natural weighting. We combined all of these observations to and achieved an r.m.s. noise of $12~\mu$Jy/beam and resolution of $1.96\arcsec \times 1.71 \arcsec$. While still suffering from the missing short spacings this image showed excellent dynamic range and filamentary structure along the western rim of \obj.

\subsection{Infrared}
The cold environment surrounding \obj\ is revealed by infrared (IR) emission. To aid in the discussion of the morphology, we made use of data from the SAGE survey of the LMC (Meixner et al. 2006) with the {\it Spitzer Space Telescope} (Werner et al. 2004). During the SAGE survey, a 7\degr$\times$7\degr\ area of the LMC was observed with the Infrared Array Camera (IRAC, Fazio et al. 2004) in the 3.6~$\mu$m, 4.5~$\mu$m, 5.8~$\mu$m, and 8~$\mu$m bands, and with the Multiband Imaging Photometer (MIPS, Rieke et al. 2004) in the 24~$\mu$m, 70~$\mu$m, and 160~$\mu$m bands. The MIPS 24~$\mu$m images provide us with a picture of the stochastically and thermally heated dust in the region of \obj\ to give an indication of the distribution of cool material. We obtained the 24~$\mu$m MIPS mosaicked, flux-calibrated (in units of MJy~sr$^{-1}$) images processed by the SAGE team from the NASA/IPAC Infrared Science Archive\footnote{See \burl{http://irsa.ipac.caltech.edu/data/SPITZER/SAGE/}}. 
The pixel sizes correspond to $4.8\arcsec$ for the 24~$\mu$m band, $\sim1.2$~pc at the LMC distance.
\section{Analysis}
\subsection{Expansion velocity of the \ha\ shell}
\label{section:ha-exp}
Both \citet{Chu1988} and \citet{Chu1997} searched for high velocity material at various regions around the shell of \obj\ using long-slit optical spectroscopy. However, in both cases, only velocities $<100$~km~s$^{-1}$ were observed, as illustrated in Fig.~\ref{figure:spectro-images} using the same spectroscopic data of \citet{Chu1997}. The \ha\ line in each of the spectro-images falls predominately in the 200--300~km~s$^{-1}$ range, consistent with the systemic velocity of the LMC \citep[250--300~km~s$^{-1}$,][]{Richter1987}. Some structure can be seen in the \ha\ line, but never with a velocity of more that 100~km~s$^{-1}$ from the centroid of the line. As noted in \citet{Chu1988} and \citet{Chu1997}, this suggests that the \ha\ shell of \obj\ is expanding at a rate consistent with an evolving superbubble rather than a supernova remnant. Note also that the presence of \nii~$\lambda6548$ and \nii~$\lambda6583$ in the spectra is indicative of radiative shocks, which occur only for low velocity shocks \citep[$\lesssim200$~km~s$^{-1}$,][]{Blair2017}.

\subsection{Measurement of the \bfield}
\label{section:bfield-calc}
As we will discuss in Sect.~\ref{mw_morphology}, the synchrotron shell of \obj\ results from a SNR that has evolved inside the superbubble. The forward shock has expanded into the hot ($\sim10^{6}$~K), rarefied medium which must have a density of $\sim10^{-3}$~cm$^{-3}$. Indeed, assuming an explosion site at the centre of \obj, HC15 found that for a current shock velocity of $\gtrsim3000$~km~s$^{-1}$ to produce synchrotron X-rays, an interior density of $\sim5\times10^{-4}$~cm$^{-3}$ is required. As shown in \citet{Weaver1977} and in recent 3D hydrodynamical simulations by \citet{Krause2018}, for young superbubbles expanding into a homogeneous environment, the interior density profile is more or less flat until very close to the supershell. Therefore, it is likely that the SNR has and continues to evolve into a relatively homogeneous medium of very low density and high temperature, accelerating VHE electrons which give rise to the filaments via synchrotron losses in the downstream region as is typical of field SNRs.

\citet{Helder2012} give an equation (their equation 26) relating the observed filament width to the postshock \bfield:

\begin{equation}
\label{eq:b_field_helder}
B_{2} \approx 26 \left(\frac{l_{\mathrm{adv}}}{1.0\times10^{18}~\mathrm{cm}}\right)^{-2/3}
\eta^{1/3}_{\mathrm{g}}\left(r_{4} - \frac{1}{4}\right)^{-1/3} \mu \mathrm{G},
\end{equation}
where, the energy-dependent, $\eta_\mathrm{g}\equiv \lambda_\mathrm{mfp}/r_\mathrm{g}$,
i.e., the ratio between the particle's mean free path and the gyroradius, and $r_4$ is the shock-compression ratio in units of 4.
This equation has been derived by balancing the acceleration time scale with the synchrotron loss time scale
and, for that reason, also contains the shock compression factor \citep{Helder2012}. However, as shown in \citet{Helder2012} this
condition is very similar to the condition that the advection length scale and diffusion length scale are equal, which leads
to $l=\sqrt{D_2\tau_\mathrm{syn}}$, which gives a nearly identical expression \citep[c.f.][]{Voelk2005,Vink2006,Rettig2012,Ressler2014}.
Similar to X-ray synchrotron spectra of various young supernova remnants, 
the X-ray synchrotron spectra of 30 Dor C \citep[$\Gamma=2-3$,][see below also]{Bamba2004} are steeper than
expected for diffusive shock acceleration and indicates that the spectra are indeed steepened due to radiative losses and must be
near the spectral cut-off, justifying the use
of Eq.~\ref{eq:b_field_helder}. Moreover, $\eta_\mathrm{g}$ must be close to unity as for shock velocities below $5000~\mathrm{km\ s^{-1}}$ X-ray synchrotron radiation can only be produced for $\eta \lesssim 10$ \citep[e.g.][]{Zirakashvili2007,Helder2012}.
In Appendix~A we discuss how our magnetic field estimates might be affected if some of the underlying assumptions are not valid.

\begin{figure*}[!ht]
\begin{center}
\resizebox{\hsize}{!}{\includegraphics[trim= 0cm 0cm 0cm 0cm, clip=true, angle=0]{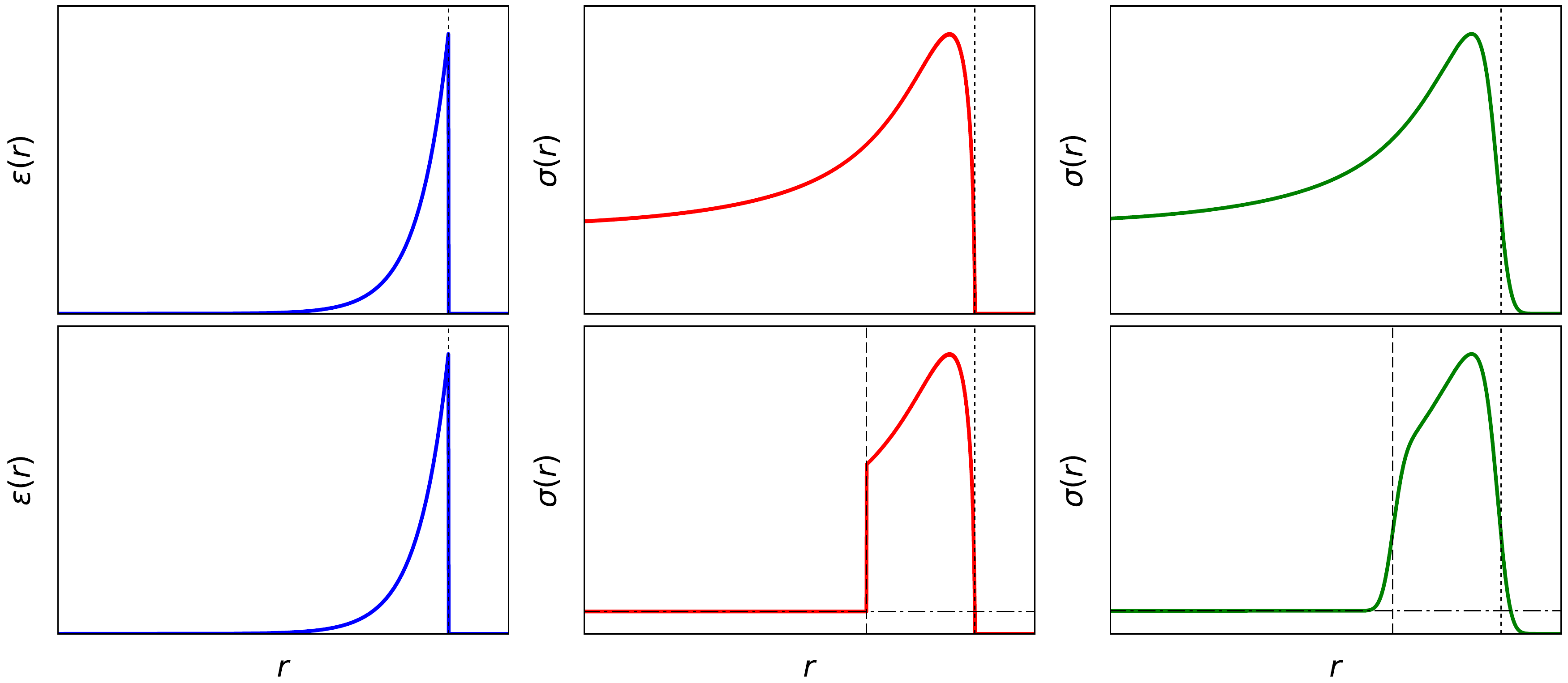}}
\caption{Illustration of projected and convolved volumetric emissivity profiles fitted to the segment profiles. {\it Top-row:} Postshock volumetric emissivity profile with radius $R$ (dotted lines) in Eq.~\ref{eq:postshock_profile} (left), projected emissivity profile using Eq.~\ref{eq:abel} (middle), and convolved with the PSF (right). {\it Bottom-row:} Same as top but for the projected `cap' model in Eq.~\ref{eq:cap_profile}. The same postshock volumetric emissivity profile is used (left). The projected emissivity profile is modified to fall to a fitted background value $b$ (dash-dot lines) below $r_{f}$ which is a fraction $r_{c}$ of the shell radius $R$, $r<r_{f}=r_{c} R$ (middle, dashed lines). This was then convolved with the PSF (right).}
\label{figure:profile_examples}
\end{center}
\end{figure*}

The actual width of the synchrotron emitting shell is somewhat larger than the advection/diffusion length scale
$l_{\mathrm{obs}} \approx \sqrt{2}l_{\mathrm{adv}}$, with the factor $\sqrt{2}$ taking into account the combination of diffusion and advection. 
However, we cannot simply measure $l_{\mathrm{obs}}$ from the \chandra\ images since the surface brightness profile we observe is the projection of the volumetric emissivity profile onto the plane of the sky. As described in \citet{Willingale1996}, assuming a spherically symmetric shell and that the shell plasma is optically thin, the surface brightness $\sigma(r_p)$ at radius $r_{p}$ can be determined from the volumetric emissivity profile $\epsilon(r)$ using the forward Abel transform:

\begin{equation}
\label{eq:abel}
\sigma (r_{p}) = 2\int^{R}_{r_{p}} \frac{\epsilon(r)r}{(r^{2}-r^{2}_{p})^{1/2}} dr
\end{equation}

\noindent where $R$ is the radius of the shell. We applied the forward Abel transform using the PyAbel Python package\footnote{See \burl{https://github.com/PyAbel/PyAbel} for code and references}. For the volumetric emissivity $\epsilon(r)$, we assumed a simple model of an instantaneous rise at $R$, followed by an exponential fall-off in the postshock region:

\begin{equation}
\label{eq:postshock_profile}
\epsilon(r)= 
\begin{cases}
A~\mathrm{exp}(-(r-R)/l_{\mathrm{obs}}) + b_{u},& r<R\\
b_{u},              & r>R
\end{cases}
\end{equation}

\noindent where $A$ is a normalisation factor (illustrated in Fig.~\ref{figure:profile_examples}, top-row), and $b_{u}$ is the upstream background level. 

The observed profile is also subject to smearing by the telescope's PSF. 
Since the \chandra\ PSF varies as a function of position and energy, we must allow for this variation when convolving $\sigma(r_p)$ with the PSF. We did this by calculating a monoenergetic PSF, set at 2.5~keV to be in the 1.5--8~keV range used for the profiles (see below), at the centre of each radial profile bin in each exposure using the MARX ray-tracing software\footnote{see \burl{http://space.mit.edu/CXC/MARX/}} \citep{Davis2012}. These PSFs were then weighted according to the exposure time of each exposure, added, and normalised. For each bin, we fitted the resulting PSF with a 2D Gaussian model and extracted a 1D profile across this model at the same position angle as the radial profile extracted from the \chandra\ data. Fitting this with a 1D Gaussian model provided an approximate PSF width at each position along the profile. These widths were used to perform a variable width convolution of the model profiles using the Varconvolve Python package\footnote{see \burl{https://github.com/sheliak/varconvolve}}.

\begin{figure}[!ht]
\begin{center}
\resizebox{\hsize}{!}{\includegraphics[trim= 0cm 0cm 0cm 0cm, clip=true, angle=0]{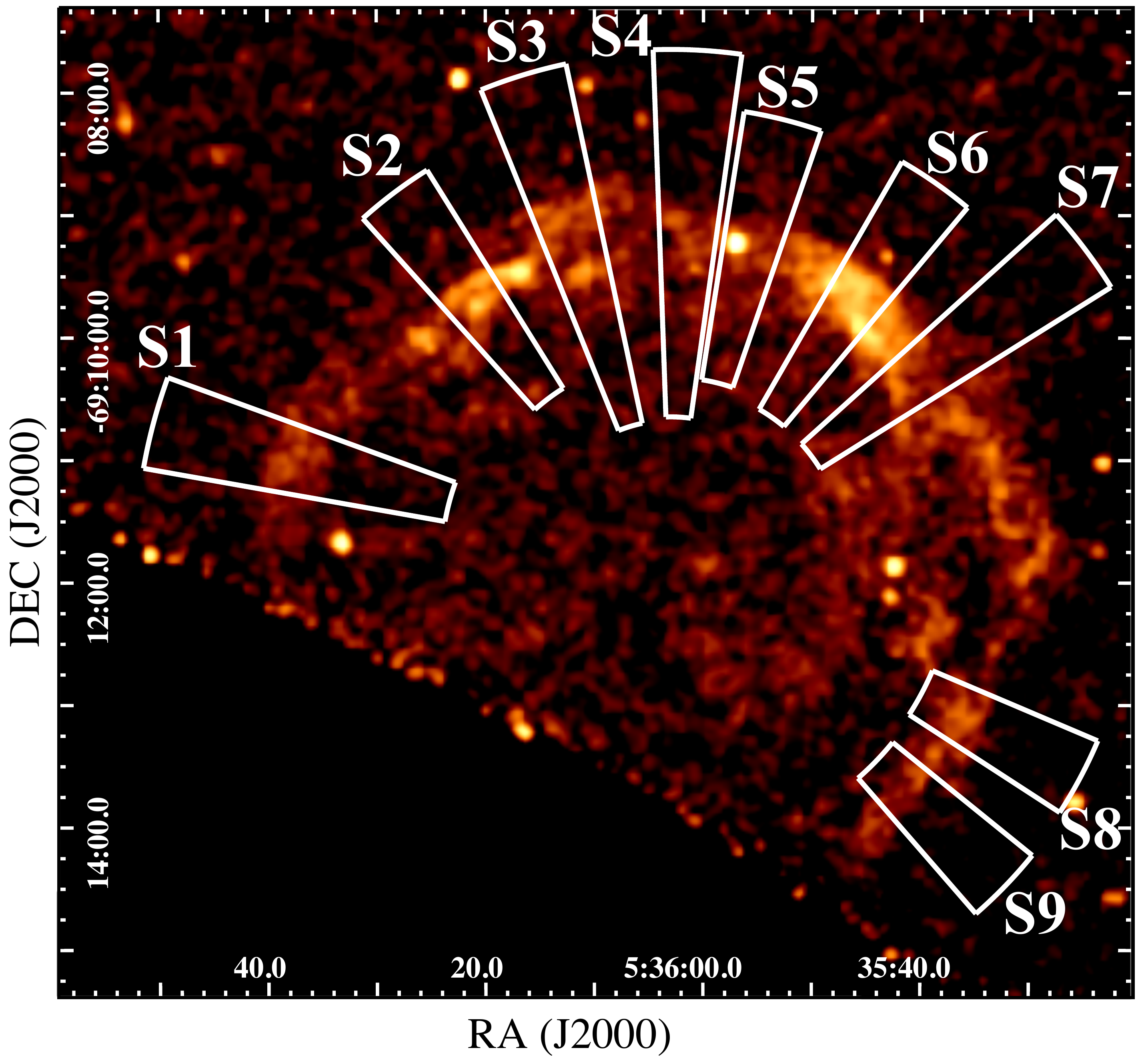}}
\caption{Segments for profile fitting overlaid on the 1.5--8~keV exposure corrected image. The image has been smoothed using a $3\sigma$ Gaussian kernel.}
\label{figure:profile_segments}
\end{center}
\end{figure}

To extract radial profiles from the data we defined sectors, centred on the \obj\ shell centre, taken from SY17, which are shown in Fig.~\ref{figure:profile_segments}. We masked the point sources, listed in Table~\ref{table:chandra_sources}, and interior enhancements that contaminated the `clean' shell profiles such as filaments projected on the interior. We extracted profiles from the combined 1.5--8~keV exposure corrected, count rate image. We set our bin sizes so that each bin had a signal-to-noise ratio $>5$ (listed in Tables~\ref{table:bfield_estimates_postshock} and \ref{table:bfield_estimates_cap}), but we omitted bins towards the centre of the shell because of low count rates and statistics. Sectors S8 and S9 were the only sectors in our sample which crossed the chip gap between the back-illuminated ACIS-S3 and front-illuminated ACIS-S4 chips. Because of the variation in sensitivity across the gap, we decided to cut the sectors and only consider the profile bins on the ACIS-S4 chip, where the brightest shell emission resides (see Fig.~\ref{figure:profile_segments}). We did not define sectors in the region between sectors S7 and S8 as the brightest emission in this region falls along the chip gap. We determined the integrated photon flux in each bin and normalised for the bin area to give radial profiles in surface brightness in units of counts~cm$^{-2}$~s$^{-1}$~arcsec$^{-2}$. 

To demonstrate that the $\Gamma$ values are indeed steep, must be
near the spectral cut-off and that we could apply Eq.~\ref{eq:b_field_helder} to determine the \bfield\ strength, we extracted and fitted spectra from each of the sectors to determine their  $\Gamma$. However we were somewhat limited by the number of counts in the majority of sectors, resulting poorly constrained values of $\Gamma$. However, much deeper observations and analyses reported in the literature do provide a good indication the values of $\Gamma$ around the shell. We used the studies of KS15 and the more recent \citet{Babazaki2018} with \xmm\ to determine indicative values of $\Gamma$ in our sectors. This resulted in $\Gamma \approx$ 2.7, 2.5, 2.6, 2.4, 2.4, 2.3, 2.4, 2.5, and 2.5 for sectors S1--S9, respectively, with a typical error of $\lesssim0.1$
.

\begin{figure*}[!h]
\begin{center}
\resizebox{\hsize}{!}{\includegraphics[trim= 0cm 0cm 0cm 0cm, clip=true, angle=0]{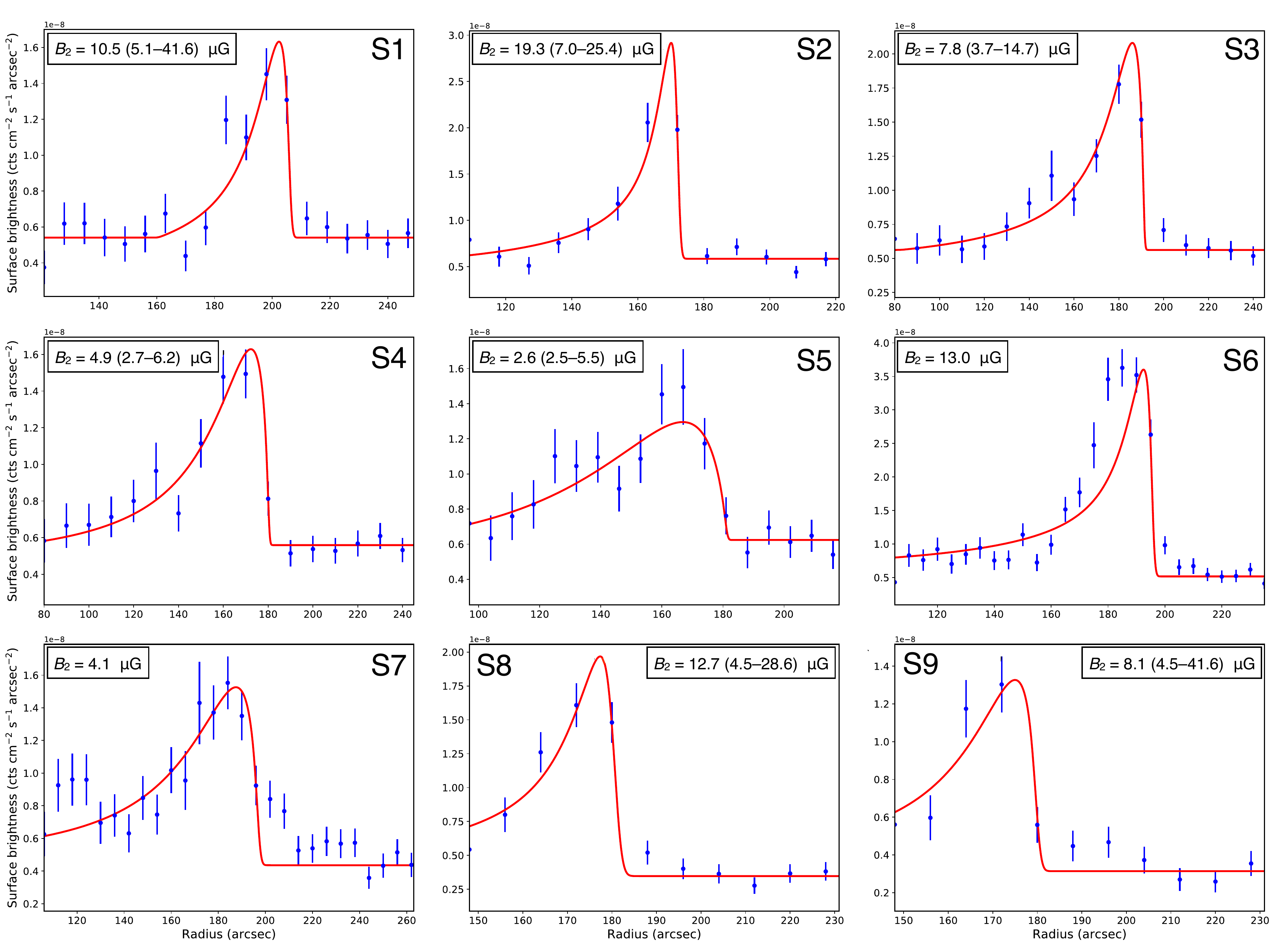}}
\caption{Radial profiles from sectors around \obj\ fitted with the postshock model described in Eq.~\ref{eq:postshock_profile}. The best fit results are given in Table~\ref{table:bfield_estimates_postshock}, with the determined $B$-fields indicated in the panels.}
\label{figure:postshock_profiles}
\end{center}
\end{figure*}

We fitted the projected and convolved volumetric emissivity profiles to the observed radial profiles allowing the values of $A$, $R$, and $l_{\mathrm{obs}}$ to vary. We determined the best-fit using $\chi^{2}$ minimisation and estimated 90\% confidence intervals for those fits with reduced-$\chi^{2}$ less than 2 ($\chi^{2}_{\nu} < 2$). This postshock model provided a good fit to most of our sectors, the results of which are given in Table~\ref{table:bfield_estimates_postshock} with the profiles and best-fit models shown in Fig.~\ref{figure:postshock_profiles}.

\begin{figure}[!h]
\begin{center}
\resizebox{\hsize}{!}{\includegraphics[trim= 0cm 0cm 0cm 0cm, clip=true, angle=0]{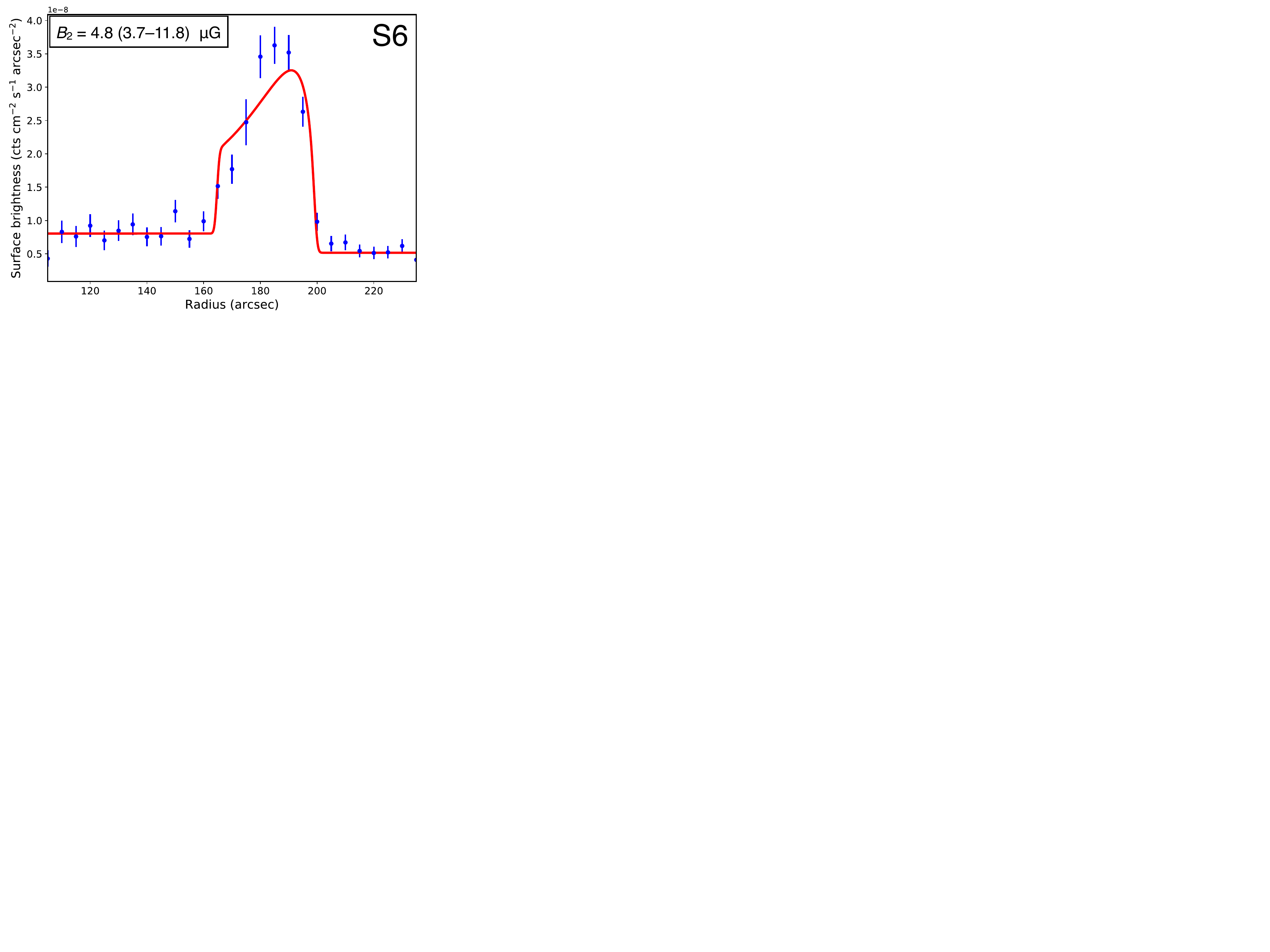}}
\caption{Radial profile from sector S6 fitted with the cap model. The best fit results are given in Table~\ref{table:bfield_estimates_cap}, with the determined $B$-field indicated.}
\label{figure:cap_profiles}
\end{center}
\end{figure}

The postshock model did not provide a good fit to sectors S6 and S7, each giving a fit statistic of $\chi^{2}_{\nu} > 2$. The S6 sector profile appeared to fall-off faster than would be expected from the postshock model. In an attempt to account for the shape of the observed profile, we modified the projected postshock profile to fall to $b_{u}$ at a fraction of the shell radius given by $r_{c} = r_{f}/ R$, essentially modelling a spherical cap of emission. However, simply allowing the profile to fall to $b_{u}$ did not account for the flux observed in the inner most bins. This is somewhat expected as the interior 1.5--8~keV flux is much brighter and has more structure in the NW quadrant than anywhere else in \obj\ (e.g., KS15). Therefore, we included a second, interior background term $b_{i}$ (which represents $b_{u}$ plus the interior flux level) and allowed the profile to fall to $b_{i}$ in the innermost bins \citep[e.g.,][]{Ressler2014}. Therefore, we applied the model:

\begin{equation}
\label{eq:cap_profile}
\sigma(r)= 
\begin{cases}
b_{i}, & r<r_{f}=r_{c} R\\
2\int^{R}_{r_{p}} \frac{\epsilon(r)r}{(r^{2}-r^{2}_{p})^{1/2}} dr,& r>r_{f}=r_{c} R\\
\end{cases}
\end{equation}

\noindent and allowed $r_{c}$ to vary in the fits. This model is illustrated in Fig.~\ref{figure:profile_examples}, bottom-row. This provided a better fit to sector S6, the results of which are given in Table~\ref{table:bfield_estimates_cap} and the profile and best-fit model shown in Fig.~\ref{figure:cap_profiles}. The cap model did not provide an acceptable fit for sector S7. We suspect that both the postshock and cap models fail to account for the profile because the shell in this sector is either not spherically symmetric, rendering the Abel transform invalid, there is too much interior structure to see a `clean' shell profile, there is an additional source of synchrotron X-rays in addition to those from the postshock region, or because there is an apparent `pre-rise' of the X-ray flux $~10\arcsec$ ahead of the main filament, which could be a faster part of the shell seen in projection.

\begin{table*}
\caption{Postshock model (see Eq.~\ref{eq:postshock_profile}) fits and \bfield\ estimates in \obj. The numbers in parentheses correspond to the 90\% confidence intervals of the fit parameters. Confidence intervals are only given for fits with $\chi^{2}_{\nu} < 2$.}
\begin{center}
\label{table:bfield_estimates_postshock}
{\renewcommand{\arraystretch}{1.3}
\begin{tabular}{lllllll}
\hline
\hline
Sector & binning & $R$ & $l_{\mathrm{obs}}$ & $l_{\mathrm{obs}}/R$ & $\chi^{2}_{\nu}$ & $B_{2}$ \\
 & (arcsec) & (arcsec) & (arcsec) & (\%) & & ($\mu G$) \\
\hline
S1 & 7 & 206.5 (204.5--212.7) & 4.7 (1.2--9.6) & 2.3 (0.6--4.7) & $1.20_{17}$ & 10.5 (5.1--41.6) \\
S2 & 9 & 172.6 (172.6--178.5) & 2.6 (1.9--7.0) & 1.5 (1.1--4.1) & $1.36_{11}$ & 19.3 (7.0--25.4) \\
S3 & 10 & 191.5 (190.6--198.0) & 6.3 (3.3--13.3) & 3.3 (1.7--7.0) & $0.59_{15}$ & 7.9 (3.7--14.7) \\
S4 & 10 & 180.8 (180.0--182.6) & 10.1 (7.9--18.5) & 5.6 (4.3--10.2) & $1.51_{15}$ & 4.9 (2.7--6.2) \\
S5 & 7 & 182.8 (175.9--183.7) & 19.3 (9.0--20.1) & 10.6 (4.9--11.4) & $0.69_{16}$ & 2.6 (2.5--5.5) \\
S6 & 5 & 195.8 & 3.8 & 1.9 & $3.81_{25}$ & 13.0 \\
S7 & 6 & 197.6 & 11.9 & 6.0 & $2.07_{25}$ & 4.1 \\
S8 & 8 & 181.2 (180.3--188.3) & 3.9 (1.7--10.9) & 2.2 (0.9--6.0) & $1.14_{9}$ & 12.7 (4.5--28.6) \\
S9 & 8 & 180.3 (172.3--181.2) & 6.1 (1.2--10.9) & 3.4 (0.7--6.3) & $1.51_{9}$ & 8.1 (4.5--41.6) \\
\hline
\end{tabular}}
\end{center}
\end{table*}%

\begin{table*}
\caption{Cap model (see Eq.~\ref{eq:cap_profile}) fits and \bfield\ estimates in \obj. The numbers in parentheses correspond to the 90\% confidence intervals of the fit parameters. Confidence intervals are only given for fits with $\chi^{2}_{\nu} < 2$.}
\begin{center}
\label{table:bfield_estimates_cap}
{\renewcommand{\arraystretch}{1.3}
\begin{tabular}{lllllll}
\hline
\hline
Sector & $R$ & $l_{\mathrm{obs}}$ & $r_{c}$ & $l_{\mathrm{obs}}/R$  & $\chi^{2}_{\nu}$ & $B_{2}$ \\
 & (arcsec) & (arcsec) & & (\%) & & ($\mu G$) \\
\hline
S6 & 198.1 (195.7--199.6) & 10.3 (4.2--13.2) & 0.82 (0.81--0.84) & 5.2 (2.1--6.7) & $1.86_{24}$ & 4.8 (3.7--11.8) \\
\hline
\end{tabular}}
\end{center}
\end{table*}%

\section{Discussion}
\label{discussion}
\subsection{Multi-wavelength morphology}
\label{mw_morphology}
\begin{figure*}
\begin{center}
\resizebox{\hsize}{!}{\includegraphics[trim= 0cm 0cm 0cm 0cm, clip=true, angle=0]{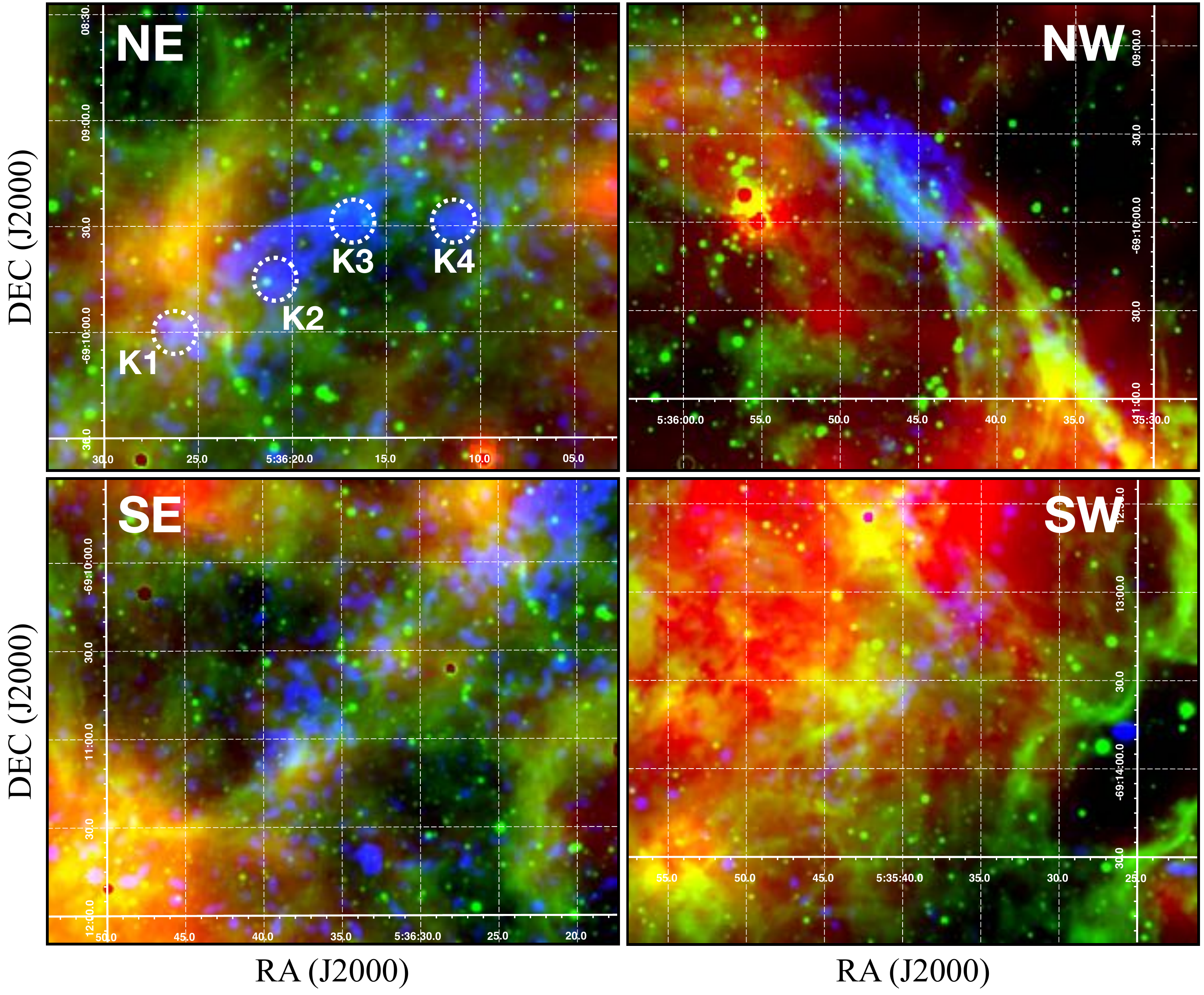}}
\caption{False-colour image of the \obj\ shell in the NE (top-left), NW (top-right), SE (bottom-left), and SW (bottom-right) with RGB = 24~$\mu$m, \ha, 1.5--8~keV. Knots K1--K4 are shown in the NE shell.}
\label{figure:multiwave_regions}
\end{center}
\end{figure*}

\begin{figure*}
\begin{center}
\resizebox{\hsize}{!}{\includegraphics[trim= 0cm 0cm 0cm 0cm, clip=true, angle=0]{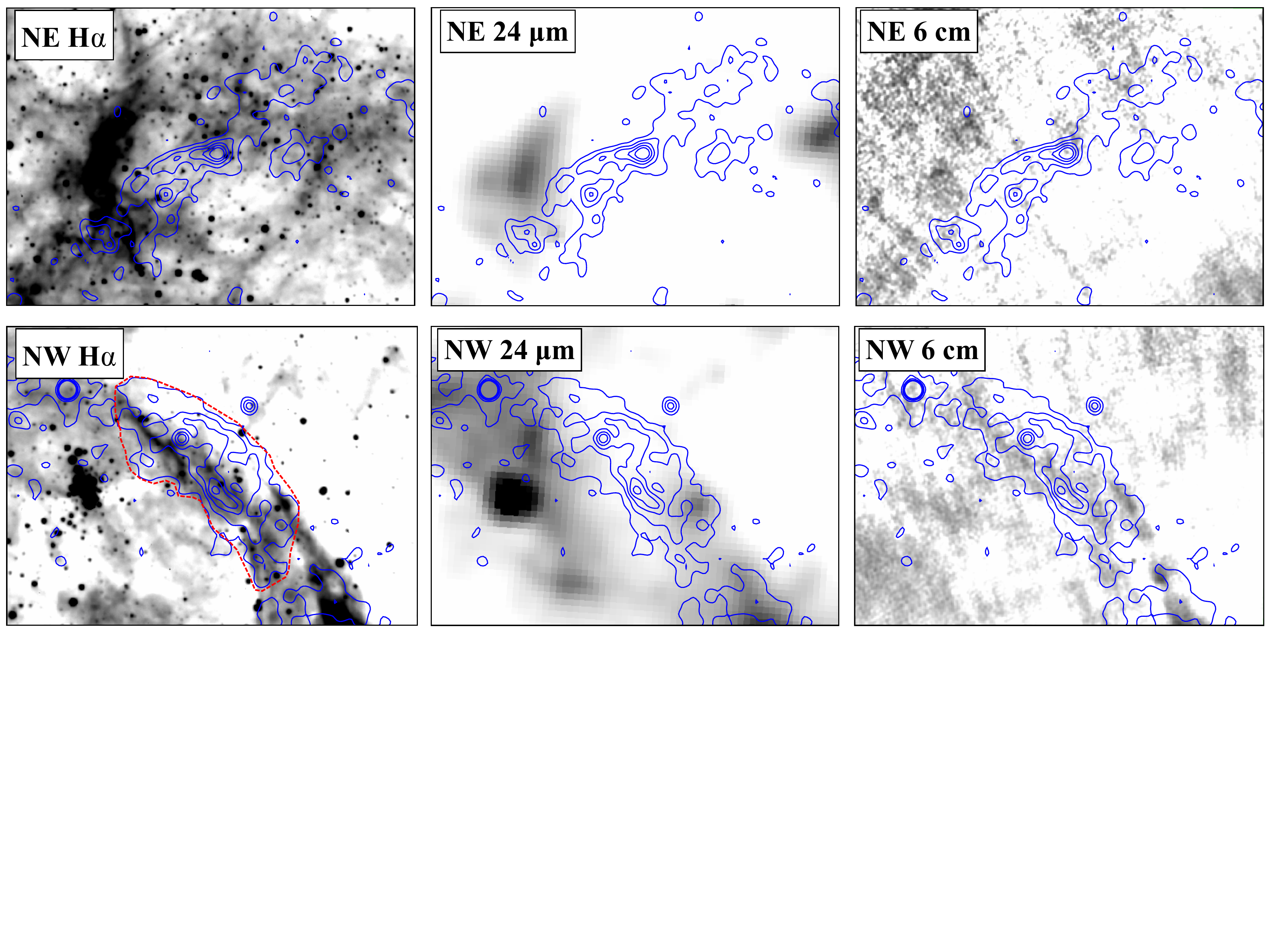}}
\caption{MCELS2 \ha\ and 24~$\mu$m images of the NE and NW shells of \obj\ with 1.5--8~keV contours overlaid. The contour levels were manually set to highlight the brightest regions of the synchrotron shell. The red region overlaid on the NW \ha\ image in bottom-left corresponds to the spectral extraction region used in Sect.~\ref{disc:synch_profiles}.}
\label{figure:multiwave_contours}
\end{center}
\end{figure*}

The new \chandra, MCELS2 \ha, and 6~cm radio continuum images provide us with the sharpest view of the brightest regions of the shell of \obj\ to date. In Fig.~\ref{figure:multiwave_regions} we show three-colour images comprising 24~$\mu$m, \ha\, and 1.5--8~keV for RGB, for the northeast (NE, top-left), northwest (NW, top-right), southeast (SE, bottom-left), and southwest (SW, bottom-right). The 24~$\mu$m is included to highlight colder material in and around the shell. Interestingly, comparing the X-ray and \ha\ emission in the NE, NW, and SE suggests that the X-ray and \ha\ shells are not correlated, as was suggested by KS15 using poorer resolution \xmm\ and MCELS data. Rather, the synchrotron X-rays fill gaps in the \ha\ shell in some regions (NE, NW) and are located ahead of the \ha\ shell in others (NW, SE). There are notable morphological consistencies in the NE and NW regions in particular with bright X-ray filaments delineating the edges of filaments in the \ha\ shell, further highlighted in Fig.~\ref{figure:multiwave_contours}. There is also little correlation between the colder material revealed in 24~$\mu$m and the synchrotron X-ray shell. Rather, the X-rays appear brighter in regions with comparatively lower levels of infrared emission. 

\begin{figure*}
\begin{center}
\resizebox{\hsize}{!}{\includegraphics[trim= 0cm 0cm 0cm 0cm, clip=true, angle=0]{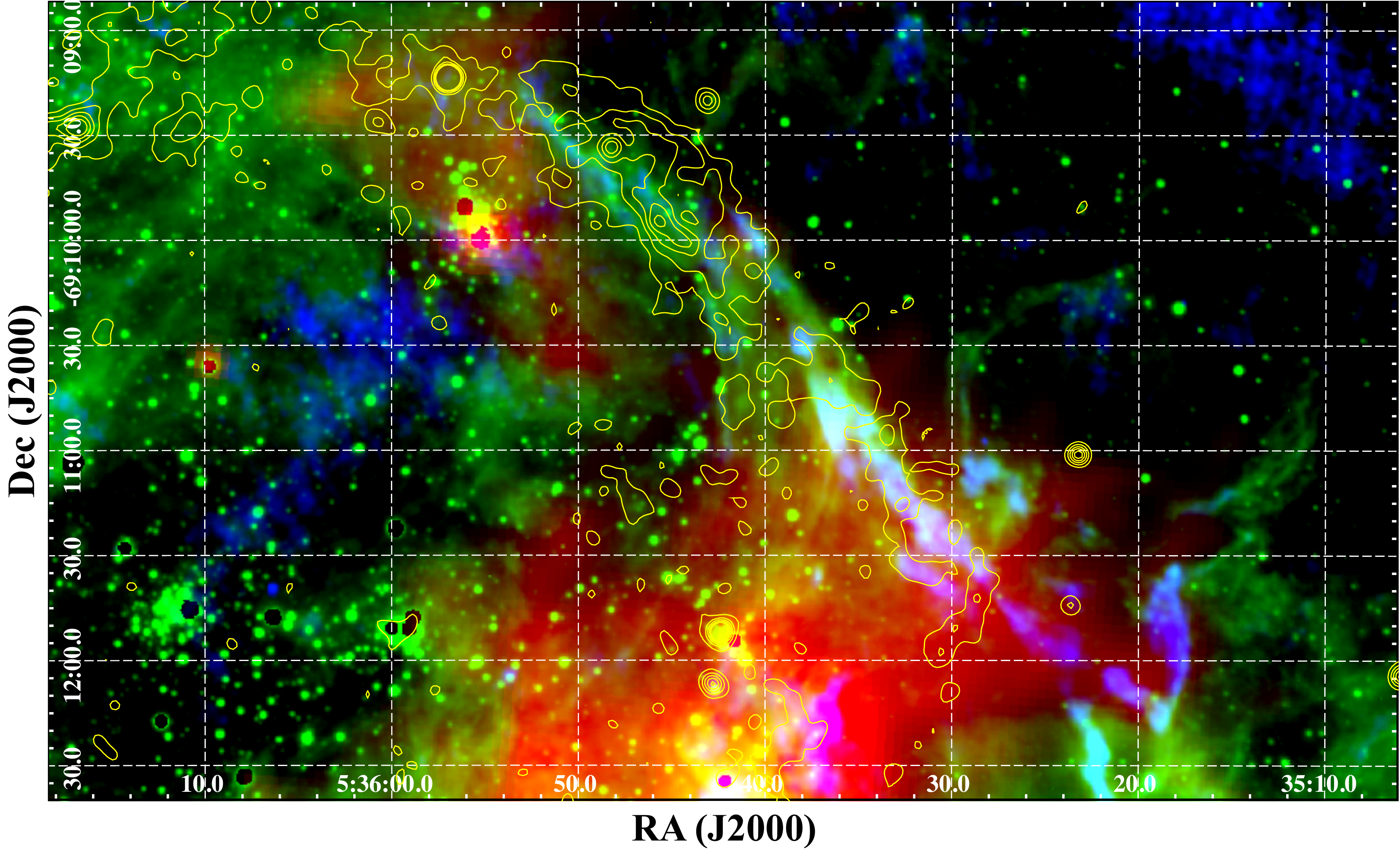}}
\caption{False-colour image of the \obj\ shell in the NW with RGB = 24~$\mu$m, \ha, 6~cm radio continuum. The contours shown in yellow are the 1.5--8~keV contours from Fig.~\ref{figure:multiwave_contours}. The extended 6~cm emission seen interior to the NW shell and in the top-right of the image is due to the edge of the primary beam.}
\label{figure:multiwave_radio}
\end{center}
\end{figure*}

We show the high spatial resolution 6~cm radio continuum image along with the 24~$\mu$m and MCELS2 \ha\ in an RGB image in Fig.~\ref{figure:multiwave_radio}. The radio continuum data bear a striking similarity to the \ha\ emission, particularly along the filaments of the NW shell. Indeed, the only deviation along the brightest filament is in regions where foreground dust, revealed by the 24~$\mu$m emission, absorbs the \ha\ emission. Therefore, the radio continuum must be thermal in origin and have little or no relation to the expanding X-ray synchrotron shell, also seen in other LMC superbubbles such as LHA~120-N~70  \citep{DeHorta2014}.

As discussed in Sect.~\ref{section:ha-exp}, the expansion velocity of the \ha\ shell is $<100$~km~s$^{-1}$, much less than the expansion velocity of the interior SNR required to explain the synchrotron X-rays \citep[$\gtrsim3000$~km~s$^{-1}$,][]{Zirakashvili2007}, as seen, for example, in the prototypical synchrotron-dominated \object{SNR~RX~J1713.6-3946} \citep{Acero2017}. The observed anti-correlation of the X-ray and \ha\ shells does suggest a resolution to this expansion velocity conflict. It is possible that the SNR responsible for the synchrotron X-ray shell has reached the \ha\ shell and has stalled in some regions, but continues through gaps in the \ha\ shell in others, and explains why the \ha\ shell is expanding at a rate typical of SBs whereas the SNR shell maintains the $\gtrsim3000$~km~s$^{-1}$ necessary to produce X-ray synchrotron emission. In addition, the bright 24~$\mu$m emission in the north, located between the bright regions of the X-ray synchrotron shell in the NE and NW (see Fig.~\ref{figure:multiwave_regions}, top-right), corresponds to a region of high radio polarisation (KS15, Fig.~7). This also supports the scenario that the expanding shock has met and compressed denser material in the north but continues to expand rapidly in the NE and NW.

The anti-correlation between \ha\ and X-ray synchrotron emission is reminiscent of a similar anti-correlation in
\object{Tycho's SNR} and RCW 86. For Tycho's SNR the non-radiative \ha\ filaments are more concentrated on the eastern side,
whereas the synchrotron filaments are on the western side \citep{Hwang2002}. It has been speculated that this anti-correlation
in caused by the damping of Alfv\'en waves if the neutral fraction is too high, which then leads to a suppression of turbulence necessary for the fast particle acceleration that gives rise to X-ray synchrotron emission.
In \object{RCW 86} a similar mechanism may also be at work, but it is more likely that the anti-correlation is caused by
large velocity gradients along the shock wave \citep{Vink2006,Helder2013}. The contrast in velocity in RCW~86 is very large, which has been attributed to the fact that this remnant evolves in a wind-blown cavity \citep{Vink2006,Williams2011,Broersen2014}. In the SW of the remnant shock velocities are lower than 500~km\,s$^{-1}$,
whereas in the NE, at the location of X-ray synchrotron emission, the shock velocity has recently been measured to
be $\sim$3000~km\,s$^{-1}$ \citep{Yamaguchi2016}. In the same region there are patches of \ha\ emission, but these appear to
be slower than the X-ray synchrotron filaments, with a mean velocity $\sim1200$~km\,s$^{-1}$ \citep{Helder2013}.

The anti-correlation in \obj, with its measured velocity contrasts, seems therefore to be a result 
of the same processes as in RCW 86, but even more extremely so. If the X-ray synchrotron filaments are the result of
a single supernova explosion going off in the extremely tenuous interior of a superbubble, the extreme velocity contrast
may be caused by density gradients and the fact that the shock radius is so much larger, $\sim50$~pc \citep[e.g.,][]{Sano2017}, that most
of the shock energy has been distributed over a large shock area, making it more sensitive to density gradients.

The difference in X-ray morphology between \obj\ and other superbubbles has been discussed by various authors (e.g., BU04, KS15). The rim-brightened morphology and hard X-rays of \obj\ contrasts the more `typical' picture of a superbubble with a centrally-filled soft X-ray morphology, such as \object{N~70} \citep{Zhang2014}. However, the optical and radio properties of \obj\ are consistent with other LMC superbubbles. The anti-correlation between synchrotron X-ray and \ha\ shell presented in this work supports that \obj\ is similar to other superbubbles but only special in that we are seeing a recent SN in the interior (see also discussions in BU04, HC15, for examples). 

\subsection{\bfield\ for hadronic models}
\label{sect:hadronic_bfield}
\begin{figure}[!t]
\begin{center}
\resizebox{\hsize}{!}{\includegraphics[trim= 0cm 1.5cm 0cm 0cm, clip=true, angle=0]{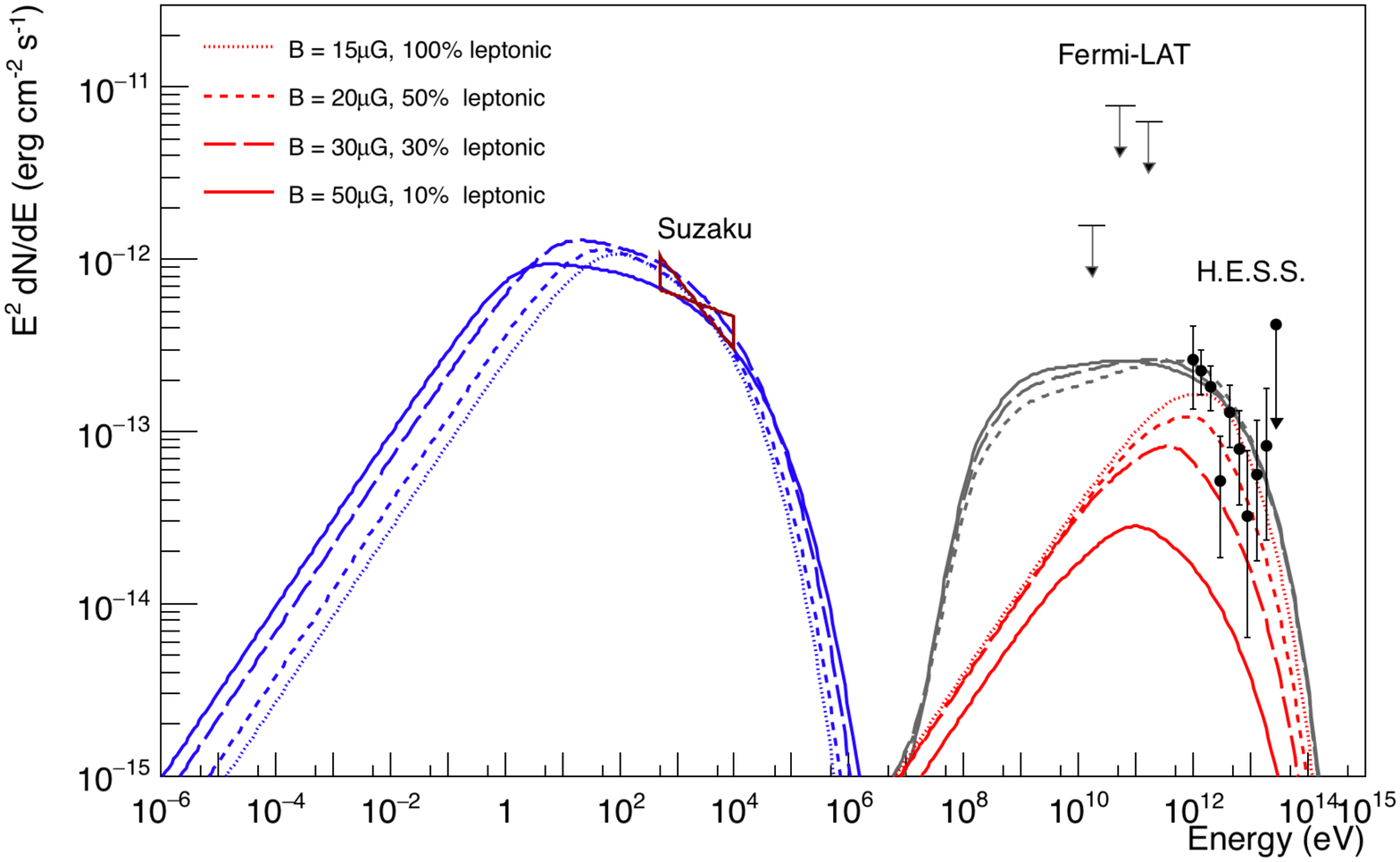}}
\caption{Three hybrid models and the pure leptonic model from HC15 applied to the SED of \obj\ described in HC15. The blue lines show the synchrotron component, the red lines show the  contribution of IC in the TeV domain, while the grey lines show the IC+hadronic component. The solid lines are for energy in protons $E_{p} = 1.50\times10^{50}$~erg, energy in electrons $E_{e} = 2.25\times10^{48}$~erg, $B = 50~\mu\mathrm{G}$, long-dashed lines for $E_{p} = 1.25\times10^{50}$~erg, $E_{e} = 3.40\times10^{48}$~erg, $B = 30~\mu\mathrm{G}$, short-dashed lines for	$E_{p} = 1.00\times10^{50}$~erg, $E_{e} = 3.50\times10^{48}$~erg, $B = 20~\mu\mathrm{G}$, and the dotted line is the pure leptonic model from HC15 with $B \sim15~\mu\mathrm{G}$. An increasing \bfield\ is required to account for the X-ray emission when the energy in protons, relative to electrons, is increased.
}
\label{figure:seds}
\end{center}
\end{figure}

To estimate a lower limit for the \bfield\ for a hadronic dominant TeV emission, we ran a set of hybrid models and compared these to the spectral energy distribution of \obj\ shown in HC15 (their Fig.~3) to illustrate how an increasing hadronic contribution to the TeV emission also requires an increasing \bfield. We show these models in  Fig.~\ref{figure:seds}, along with the purely leptonic model from HC15. As the energy in protons, relative to electrons, is increased, the \bfield\ required to fit the synchrotron X-rays also increases. For the hybrid model with completely dominant hadronic TeV emission ($>$90\%), a \bfield\ of $\gtrsim50~\mu$G is needed to account for the X-ray emission. The model with a 50-50 contribution to the TeV emission requires a \bfield\ of $\sim20~\mu$G.

\subsection{Synchrotron profiles}
\label{disc:synch_profiles}
In Sect.~\ref{section:bfield-calc}, we described the extraction of synchrotron emission profiles from various sectors around the shell and their modelling with a radial profile as typically seen from SNRs, i.e., an instantaneous rise at shell radius $R$, followed by an exponential fall-off in the postshock region and assuming the shell is spherically symmetric. In almost all sectors (S1, S2, S3, S4, S5, S8, and S9), this model provided a good fit to the radial profiles. The fact that the profiles fit this SNR model in most regions of \obj, and the anti-correlation between \ha/ 24~$\mu$m emission (Fig.~\ref{figure:multiwave_regions}) and X-ray synchrotron emission
argue against the interpretation of SY17 that the synchrotron X-rays originate in the shock-cloud interaction regions. If this were the case, the observed profiles should be the sum of a multitude of very narrow synchrotron filaments in the various shock-cloud interaction regions, and there is no reason to expect that this would give rise to the SNR-type volumetric emissivity profile that provides a good fit to the data. 

However, there are two sectors whose radial profiles are not well-fitted by the SNR model, i.e., sectors S6 and S7. Interestingly, these sectors cross the brightest region of the synchrotron shell which is correlated with the MC4 molecular cloud identified by SY17. Therefore, it is possible that some or all of the synchrotron emission in the brightest region could be due to VHE electrons in shock-cloud interaction regions. In Sect.~\ref{section:bfield-calc} we showed that the S6 profile can be fitted using a modified `cap' model. While this does provide an acceptable fit to the data, we have no reason to expect such an emission profile in this sector. In addition, some bright X-ray knots of emission were found in the NE shell, which were masked during the radial profile extraction. The origin of these knots, which are marked in Fig.~\ref{figure:multiwave_regions} as knots K1 through K4, is unclear. If these knots resulted from enhanced emission at shock-cloud interaction regions, we might expect them, and the NW shell, to be variable on short timescales \citep{Inoue2009} such as in supernova remnants such as \object{Cas~A} on timescales of a few years \citep{Uchiyama2008}.

\begin{figure*}
\begin{center}
\resizebox{\hsize}{!}{\includegraphics[trim= 0cm 0cm 0cm 0cm, clip=true, angle=0]{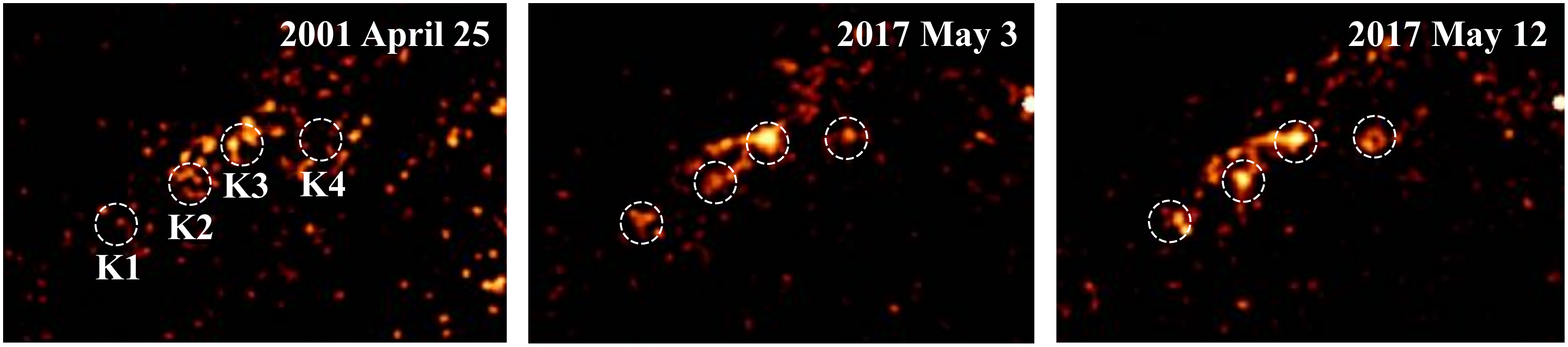}}
\caption{\chandra\ 1.5--8~keV flux corrected images of the NE shell over three epochs, which are indicated in the panels. Knots K1--K4 are shown on the plots. Each image has been smoothed using a 3$\sigma$ kernel to better show the diffuse structure. The colour scales in each panel are the same.}
\label{figure:varibility}
\end{center}
\end{figure*}

To search for variability, we used both epochs of our new \chandra\ observations (2017 May 3 and 2017 May 12) and the only other observation to cover the northern shell, taken in 2001 April (ObsID 1044, $\sim$18 ks, PI: G. Garmire) which were reported in BU04. We processed this dataset as described in Sect.~\ref{section:chandra-obs}. The target of ObsID~1044 was \object{SN1987A}, and, therefore, the NW and NE shells are located $\gtrsim$7' off-axis, resulting in lower sensitivity and a degradation of the PSF. Flux-corrected and smoothed 1.5--8~keV images of the NW and NE shells in each epoch are shown in Fig.~\ref{figure:varibility}, along with the positions of the knots. We used the observed counts to determine the photon flux and error in each knot region for each epoch. We found for Obs. ID 1044 that the comparatively short exposure time of $\sim$18 ks, coupled with the off-axis location of the northern shell resulted in a small number of counts (<20) per knot in the 1.5--8~keV range, prohibiting a robust variability study of the knots. In addition, the brightest region of the NW shell  was located on an ACIS-S chip gap during ObsID~1044 so we could not reliably compare counts between the 2001 and 2017 epochs.

Interestingly, in the course of our variability study, we found that the NW shell region appeared to vary between the two 2017 epochs, i.e., on a timescale of nine days. The increase from $51.49~(\pm 2.40)
\times 10^{-6}$~phot~cm$^{-2}$~s$^{-1}$ on 2017 May 3 to $64.24~(\pm 2.52)
\times 10^{-6}$~phot~cm$^{-2}$~s$^{-1}$ on 2017 May 12 corresponds to a $\gtrsim10$~\% increase in flux, which is rather puzzling as synchrotron variability is not expected on such short timescales. Fluxes extracted from the other bright synchrotron region in the NE shell, located $\sim2\arcmin$ away and also on the ACIS-S3 chip, showed no evidence of variability. 

\begin{figure}[!t]
\begin{center}
\resizebox{\hsize}{!}{\includegraphics[trim= 0cm 0cm 0cm 0cm, clip=true, angle=0]{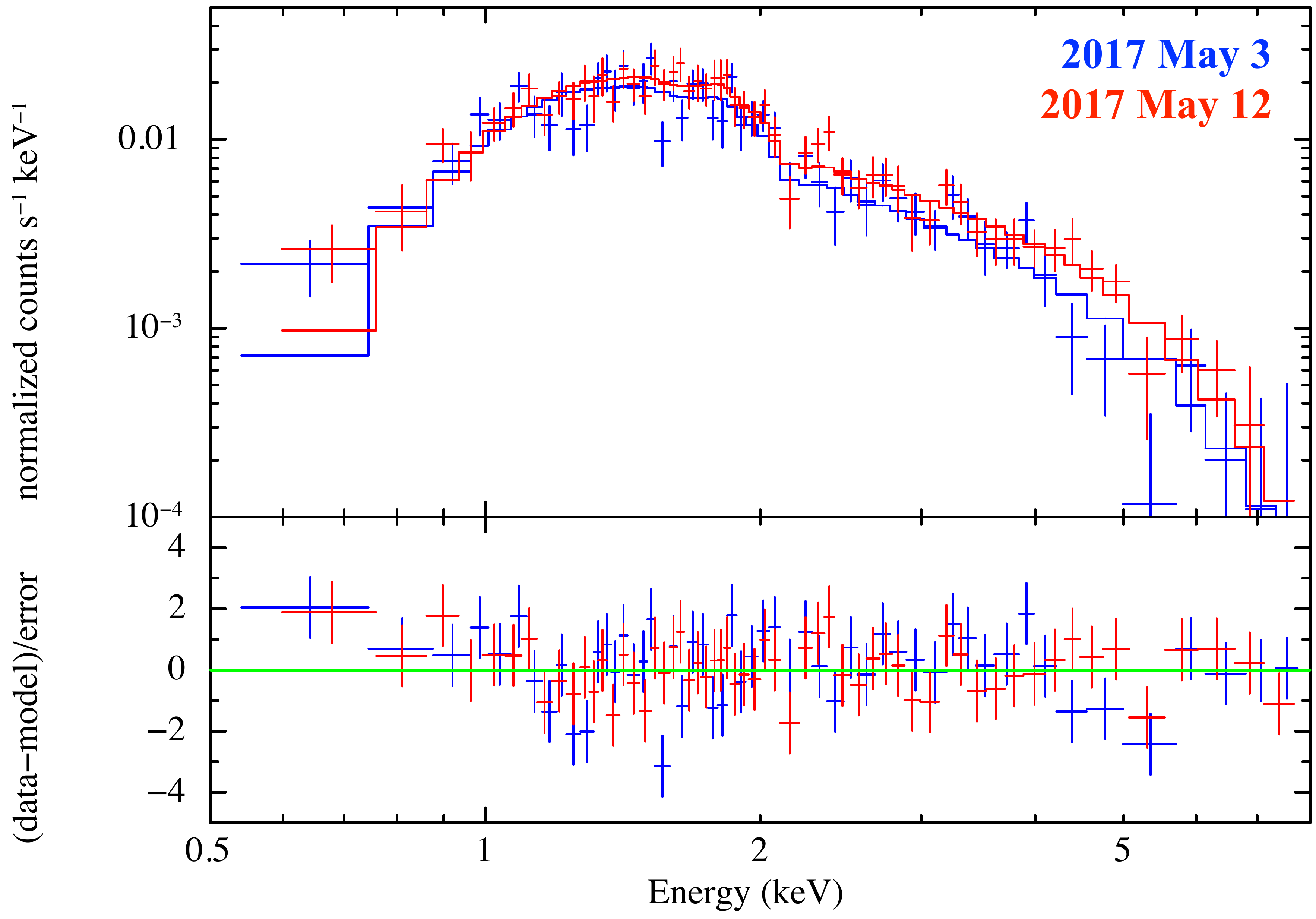}}
\caption{Spectra from the NW shell fitted with an absorbed power-law model. The spectrum from 2017 May~3 is shown in blue and that from 2017 May~12 in red.}
\label{figure:NW-spectra}
\end{center}
\end{figure}

We further assessed the increase in flux and whether it was accompanied by a change in spectral shape, using spectral analysis. We extracted source (indicated in Fig.~\ref{figure:multiwave_contours}, bottom-left) and background spectra for the NW shell using the CIAO task \texttt{specextract} and fitted them using XSPEC \citep{Arnaud1996} version 12.8.2p with abundance tables set to those of \citet{Wilms2000}, photoelectric absorption cross-sections set to those of \citet{Bal1992}. Detected point sources were masked. Since the spectrum of the NW shell has been found in all previous studies to be dominated by a single non-thermal component, we fitted the spectra with a power-law, absorbed by Galactic and LMC material (\texttt{phabs*vphabs*pow} in XSPEC), with the Galactic absorption fixed to $6\ \times~10^{20}$~cm$^{-2}$ \citep{Dickey1990} and the LMC absorption fixed to $1.0\ \times\ 10^{22}$~cm$^{-2}$ (KS15). We estimated fluxes and errors using the \texttt{cflux} convolution model component. Our spectral model provided a good fit in each epoch, and the data and best-fits are shown in Fig~\ref{figure:NW-spectra}. The increase in flux between 2017 May~3 and 2017 May~12 is evident in the plot and the best-fit \texttt{cflux} parameters with $F_{\rm{X, 1.5-8~keV}} = 2.19~(1.95-2.39)\times10^{-12}$~erg~cm$^2$~s$^{-1}$ for 2017 May 3 and $F_{\rm{X, 1.5-8~keV}} = 2.95~(2.69-3.16)\times10^{-12}$~erg~cm$^2$~s$^{-1}$ for 2017 May 12. The best-fit photon index ($\Gamma$) of the power-law component does decrease between the epochs, however we cannot conclude that this is in fact the case as the indices are consistent within the 90\% confidence intervals for 2017 May~3 and 2017 May~12 at $\Gamma_{2017-05-03} = 2.55~(2.41-2.70)$ and $\Gamma_{2017-05-12} = 2.32~(2.20-2.44)$, respectively.

While the determined fluxes from the NW shell at the 2017 May~3 and 2017 May~12 epochs suggest an increase in brightness of the shell of $\gtrsim$10\%, the currently available datasets cannot show that this is accompanied by a change in the spectrum. The abrupt increase in flux is very difficult to physically explain. Taking the observed widths of the S6 and S7 sectors of the synchrotron shell from Tables~\ref{table:bfield_estimates_postshock} and \ref{table:bfield_estimates_cap}, the minimum width of the shell is $\sim4\arcsec$ which corresponds to $\sim$1~pc at the LMC distance. Given this width, signal speeds faster than the speed of light would be required to explain a flux variability on timescales of days. We also assessed a possible systematic origin for the apparent flux increase. The bounds of the NW shell region are within an arcminute of the ACIS-S aimpoint. Other regions considered for variability in the NE are $2\arcmin-3\arcmin$ away but also on the S3 chip and no evidence of variability was found in these, ruling out some variation in detector background between the epochs. We also checked for variation in the NW background region, located outside the NW shell but no variation was found. As already noted, detected point sources were masked so the increase in flux is not due to a variable point source. Future deep \chandra\ observations would be required to verify if the apparent variability is real. 

\subsection{\bfield\ estimates and TeV emission mechanism}
The estimated \bfield\ in those sectors well-fitted by the SNR volumetric emissivity profile model is low with the best fit \bfield\ strengths ranging from 2.6--19.3~$\mu$G (see Table~\ref{table:bfield_estimates_postshock}). In three sectors the upper limits of the 90\% confidence intervals extend beyond $\sim30$~$\mu$G, though the estimates in these sectors are poorly constrained. Therefore, the shape of the profiles (discussed in the previous sub-section) and the determined \bfield\ strengths suggest an SNR origin, where the average downstream magnetic field strength is consistent with a compressed ISM. These low magnetic field strengths suggest a leptonic-dominated origin for the TeV $\gamma$-rays detected by HC15 from \obj.

The upcoming Cherenkov Telescope Array \citep[C.T.A.,][]{CTA2017} will provide further constraints and verification of the TeV emission mechanism in \obj. Hadronic and leptonic mechanisms predict different flux for the (10~GeV--1~TeV) band, both currently below {\it Fermi}-LAT \citep{Fermi2009,Ackermann2011} sensitivity and between {\it Fermi} and \hess\ covered energy ranges, but within the capabilities of C.T.A. An example of the prospects of C.T.A. in this regard is shown by \citet{Acero2017b} as applied to the SNR RX~J1713.7--3946. The recent \hess\ paper on RX~J1713-3946 \citep{hess2018} demonstrates the possibilities for C.T.A. observations of LMC objects. \hess\ revealed the TeV shell of RX~J1713.7--3946 in unprecedented detail, and was found to extend further than the X-ray shell. This allowed the probing of particle escape, while the GeV and TeV spectra were covered to a level of accuracy that allowed a very detailed comparison of leptonic and hadronic emission mechanisms, including a magnetic field map for the leptonic case.

\section{Conclusion}
\label{conclusion}
We have presented an analysis of new \chandra\ observations of \obj\ in the Large Magellanic Cloud, the first superbubble detected in TeV $\gamma$-rays. These observations provided the sharpest view of the synchrotron X-ray shell of \obj, allowing us to perform a detailed morphological study and estimate the \bfield\ in the superbubble, a key discriminator in assessing the dominant TeV $\gamma$-ray emission mechanism with a low $B$-field $\sim15$~$\mu$G required for a purely leptonic origin, a $B$-field of $\gtrsim50$~$\mu$G for a completely hadronic-dominated origin, or $B$-field of $\sim20$~$\mu$G for a 50-50 contribution of the leptonic and hadronic mechanisms to the TeV $\gamma$-rays.

Using the new \chandra\ data, MCELS2 \ha, and 6~cm radio continuum images we found an anti-correlation between the synchrotron X-ray and \ha/6~cm shells. In addition, we discussed how long-slit spectroscopy of various regions of the \obj\ shell has shown no evidence of the high-velocities necessary to explain the synchrotron X-rays ($\gtrsim3000$~km~s$^{-1}$). Rather the \ha\ expansion velocities are more typical of an expanding superbubble ($<100$~km~s$^{-1}$). We suggested that the SNR responsible for the synchrotron X-rays has reached the \obj\ supershell and has stalled in some regions, but continues through gaps in the shell in others. This is similar to the observed anti-correlation seen in RCW~86 \citep{Vink2006,Helder2013}, which is attributed to the SNR evolving into a wind-blown cavity and encountering density gradients, though the velocity differences between synchrotron X-ray and \ha\ shells are not as extreme. This may be a result 
of more pronounced density gradients and/or the fact that the shock radius is so much larger in \obj, meaning the shock energy has been distributed over a large area, making it more sensitive to these density gradients.

We estimated the downstream \bfield\ from the synchrotron X-ray shell. This was achieved by fitting the observed radial profile in sectors around \obj\ with a typical postshock volumetric emissivity profile projected onto the sky and convolved with the \chandra\ PSF to determine the width of the shell. From this width we determined the \bfield\ using Eq.~26 from \citet{Helder2012}. We obtained good fits to the majority of the sectors with the postshock model and found that the downstream \bfield\ was generally low, all with best fits $\lesssim 20~\mu$G, though upper confidence limits reaching $\gtrsim 30~\mu$G in three sectors where the confidence intervals were poorly constrained. This suggests that the TeV emission is likely dominated by IC emission, i.e., the leptonic scenario. Our postshock model did not provide good fits to two sectors. We found that one sector did not provide a `clean' radial profile because of interior structure, while the other could be fitted with a modified projected postshock model where the projected profile falls off abruptly below $\sim0.8$ times the shell radius, yielding a postshock \bfield\ of 4.8~(3.7--11.8)~$\mu$G which is again consistent with the leptonic TeV $\gamma$-ray mechanism. Alternatively, the observed profiles in these sectors could result from synchrotron enhancements around a shock-cloud interaction as suggested by SY17. 

\begin{acknowledgements}
The authors wish to thank the anonymous referee for their comments and suggestions which improved the paper. The authors also wish to thank Pierrick Martin whose helpful comments further improved the paper. M.S.\ acknowledges support by the Deutsche Forschungsgemeinschaft (DFG) through the Heisenberg fellowship SA 2131/3-1, the Heisenberg research grant SA 2131/4-1, and the Heisenberg professor grant SA 2131/5-1. 
\end{acknowledgements}

\bibliographystyle{aa}
\bibliography{refs.bib}

\begin{appendix}
\section{Could the electron spectrum be age-limited instead of loss-limited?}
In the main body of the text we showed that the widths of the X-ray synchrotron filaments are typically
5\arcsec\ (5.8 pc or $1.8\times 10^{19}$~cm), which implies a
magnetic field strength of $B \lesssim 15\mathrm{\mu G}$, based on Eq.~\ref{eq:b_field_helder}. This low magnetic
field implies that TeV  $\gamma$-ray emission is dominated by inverse Compton scattering of background photons,
rather than by pion production \citep[][see also Sect.~\ref{intro}]{hess2015}.
However, the magnetic field strength estimate relies on Eq.~\ref{eq:b_field_helder}, and there may be some concern that the electron maximum energy is not limited by radiative losses.

A more general equation that relates the X-ray synchrotron width to magnetic field strength 
is the advection length scale, already alluded to in the main text:
\begin{align}\label{eq:advection}
  l_\mathrm{adv}= &<\Delta v>\tau_\mathrm{syn} =  <\Delta v> \frac{9(m_\mathrm{e}c^2)^2}{4e^4c}\frac{1}{B^2E_\mathrm{e}}\approx \frac{634}{B^2E_\mathrm{e}},
\end{align}
with $\sigma_\mathrm{T}$ the Thomson cross section, $m_\mathrm{e}$ the electron rest mass, $E_\mathrm{e}$ the electron energy,
and $<\Delta v>$ the average advection velocity downstream of the shock. 
Normally we would assume
that $<\Delta v>= V_\mathrm{s}/r$, with $r=4$ the shock compression ratio. However, here we will allow
for gradients in velocity.
This equation  is generally applicable, but has the disadvantage that it depends on  $<\Delta v>$, which we do
not know, and on the typical electron energy, $E_\mathrm{e}$, at which we observe the shock. The latter can be 
estimated by using the relation between photon energy, electron energy, and magnetic field strength for
synchrotron radiation \citep{Ginzburg1965}: $E_\mathrm{ph}= 7.4 E_\mathrm{e}^2B$~keV (with $E_\mathrm{e}$ and $B$ in cgs units).
This gives \citep[see also][]{Rettig2012}:
\begin{align}\label{eq:advection}
  l_\mathrm{adv} 
  \approx& 5.5 \times 10^{18}
  \left(\frac{<\Delta v>}{1000~\mathrm{km\ s^{-1}}}\right)
\left(\frac{E_\mathrm{ph}}{1~\mathrm{keV}}\right)^{-1/2}\left(\frac{B}{10~\mathrm{\mu G}}\right)^{-3/2}~\mathrm{cm};
\end{align}
or inverted:
\begin{align}\label{eq:b_l_adv}
B \approx 31 \left(\frac{l}{1\times 10^{18}~\mathrm{cm}}\right)^{-2/3}    \left(\frac{<\Delta v>}{1000~\mathrm{km\ s^{-1}}}\right)^{2/3} \left(\frac{E_\mathrm{ph}}{1~\mathrm{keV}}\right)^{-1/3}.
\end{align}
This expression has some similarities to Eq.~\ref{eq:b_field_helder}, since that equation was also based on the advection length scale.
It also shows that for photons around 1~keV and advection speeds of $\approx 1000~\mathrm{km\ s^{-1}}$, we get a very similar magnetic field strength.
However, unlike for some young SNRs, such as Cas~A \citep{Vink2003}, we lack a measurement of the shock speed and thus
an estimate of $<\Delta v>$. So the question then is whether $<\Delta v>\approx 1000~\mathrm{km\ s^{-1}}$ is indeed a good estimate for the advection velocity in the X-ray synchrotron
filaments of \obj.

Superbubbles are expected to have expansion velocities of $30-200~\mathrm{km\ s^{-1}}$. This is consistent with the velocity information from the optical emission from 30 Dor C, which applies to
the thermal X-ray emitting shell, not the X-ray synchrotron emitting shell.
If we would assume these velocities, say  $<\Delta v>\approx 100~\mathrm{km\ s^{-1}}$,
the magnetic field estimate would come down to $B\sim 3~\mathrm{\mu G}$ or less. This would weaken the case for hadronic gamma-ray emission even more, but such a low velocity
would be inconsistent with the emission of X-ray synchrotron radiation.
A problem with much lower shock velocities is that they cannot produce X-ray synchrotron emission, as this generally requires $V_\mathrm{s}\gtrsim 3000~\mathrm{km\ s^{-1}}$  \citep[e.g.][]{Aharonian1999,Zirakashvili2007},
corresponding to $\Delta v \gtrsim 750~\mathrm{km\ s^{-1}}$.
This requirement relies on the assumption  that acceleration gains balances radiative losses, but we note that letting go of this requirement generally leads to low magnetic fields. We illustrate this by using the approximate relation between shock speed, and the radius and age of a supernova remnants:
$V_\mathrm{s}= m R/t$,
with $t$ the age of the object and $0.4 \leq m \leq 1$, with $m=0.4$ corresponding to the Sedov-Taylor solution and $m=1$ to free expansion \citep[see for example the review][]{Vink2012}.
Now the requirement that the synchrotron loss time is longer than the age of the object implies:
\begin{align}
t=\frac{mR}{V_\mathrm{s} }<& \tau_\mathrm{syn} \approx \frac{634}{B^2E_\mathrm{e}}\\\nonumber
\Rightarrow &\\\nonumber
B < &3.1 m^{-2/3}\left(\frac{E_\mathrm{ph}}{1~\mathrm{keV}}\right)^{-1/3}
\left(\frac{V_\mathrm{s}}{5000~\mathrm{km\ s^{-1}}}\right)^{2/3}
\left(\frac{R}{50~\mathrm{pc}}\right)^{-2/3}~\mathrm{\mu G}.
\end{align}
So for an object the size of \obj\ the electron spectrum has to be loss limited, or the magnetic field has to be even lower than our best estimate. But if the spectrum is loss limited
the current shock velocity has to be $V_\mathrm{s}\gtrsim 3000~\mathrm{km\ s^{-1}}$, or it has to have been that high in the recent past, i.e, less than a synchrotron loss time scale ago.

Looking at Eq.~\ref{eq:b_l_adv} we in fact see that the only way that our main conclusion, namely  that the magnetic field strength is lower than $50~\mathrm{\mu G}$, can be wrong is if
$<\Delta v> \gtrsim 3000~\mathrm{km\ s}^{-1}$, corresponding to $V_\mathrm{s} > 12000~\mathrm{km s}^{-1}$. 
Although a very young SNR could have $V_\mathrm{s}\sim 12000~\mathrm{km\ s^{-1}}$, it would not maintain that velocity for long enough to inflate to a radius of 50~pc ($\sim 6$~kyr for $V_\mathrm{s}\sim 12000~\mathrm{km\ s^{-1}}$).

Given that $<\Delta v> \approx 1000~\mathrm{km\ s}^{-1}$ is in close agreement with expectations, and that Eq.~\ref{eq:b_l_adv} provides a very similar magnetic field
estimate as  Eq.~\ref{eq:b_field_helder}, strengthens the reliability of our magnetic field estimates, and also provides  evidence that $\eta_\mathrm{g}\lesssim 10$.

It should be noted that the magnetic field may be strongly position dependent, if magnetic field damping plays an important role, but also due
to the divergent flow of the plasma.
In the context of X-ray synchrotron emission from supernova remnants the potential role of magnetic field damping was first pointed out in \citet{Pohl2005}. The idea is that the cosmic-ray induced magnetic field amplification \citep{Bell2004},
which occurs in the cosmic-ray precursor, will decay again in the downstream region. In that case the narrow widths of the X-ray filaments may not so much reflect the advection length scale (Eqs. \ref{eq:b_field_helder} and ~\ref{eq:advection}), but the typical decay length scale of the magnetic field. This idea was applied to young SNRs by \citet{Rettig2012} and
to Tycho's SNR by \citet{Tran2015}. 
We have ignored the effect in our magnetic field estimate, but we note here that our main conclusion is that the magnetic field strength in \obj\ is
lower than expected.  
Magnetic field damping  leads to observed X-ray synchrotron filament widths that are smaller than advection/synchrotron loss models. 
In general, therefore, magnetic field estimates including damping lead to lower estimates of the magnetic strengths as reported by \citet{Rettig2012, Tran2015}. 

\begin{figure}[!h]
\begin{center}
\resizebox{\hsize}{!}{\includegraphics[trim= 0cm 0cm 0cm 0cm, clip=true, angle=0]{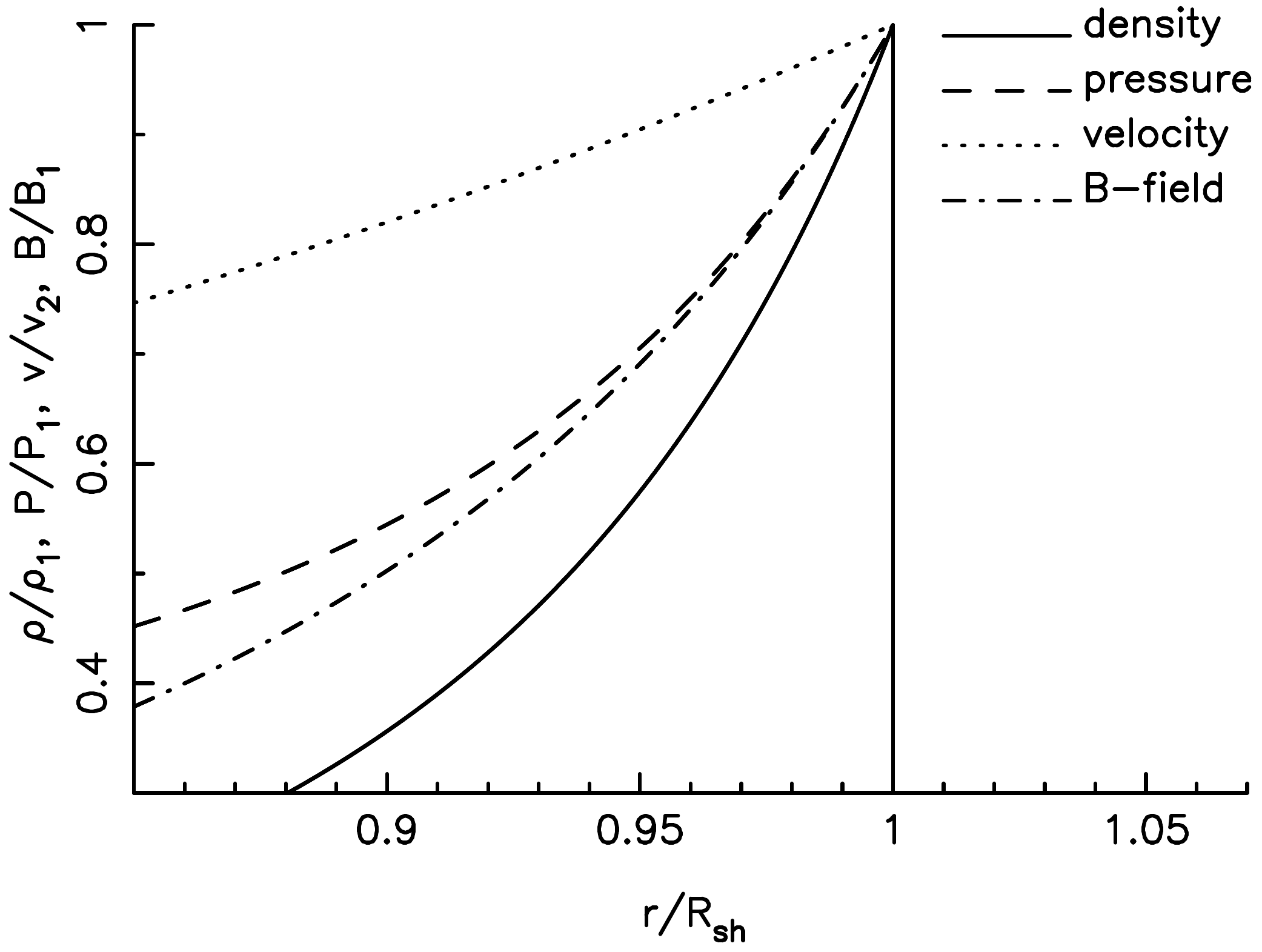}}
\caption{Profile of the density, pressure, velocity and magnetic field for the Sedov-Taylor solution.
The magnetic field profile has been estimated assuming $B\propto n^{2/3}$. All quantities scaled to
the quantities immediately downstream of the shock.
}
\label{figure:sedov}
\end{center}
\end{figure}

The effects of the divergent flow on the magnetic field profile can be estimated
based on the Sedov-Taylor model, which provides scaled densities, pressures, and velocities as a function of shock radius. For the magnetic field
we can estimate that $B\propto n^{2/3}$ (flux conservation), 
but note that the drop due to the divergence of the flow is mostly affecting the radial component of the magnetic field.
The plasma variables for the Sedov-Taylor model close to the shock front is depicted in Fig.~\ref{figure:sedov}. The average $l_\mathrm{obs}/R$ in Table~\ref{table:bfield_estimates_postshock}
is 4\%, so we see that the magnetic field may have declined by 30\% with respect to the value near the shock. A similar value was found by \citet{Zhang1996} for a supernova remnant evolving in a stellar wind bubble. 
The decline in magnetic field affects the emissivity by about 50\%, making the filaments appear a bit smaller than for a constant magnetic field.
The effects on the estimates are, therefore, qualitatively similar to magnetic field damping as both lead to lower
magnetic fields further downstream and corresponding overestimates of the magnetic field strengths. 
In light of the strong magnetic field evolution, the value of $l_\mathrm{obs}/R\approx $10\% for region S5 is somewhat surprising. However, the error on $l_\mathrm{obs}$ is rather large, and the plasma behind the shock may be different from the Sedov-Taylor solution.

In addition, note that the plasma velocity at 95\% of the shock radius is 90\% of the flow speed immediately downstream of the shock. 
The average flow velocity, $<\Delta v>$, is even closer to
the plasma velocity near the shock, as the plasma at 95\% of the shock radius was shocked at an earlier time when the shock velocity was still higher.
For an accurate estimate of $<\Delta v>$ we need a Langrangian description of the plasma, rather than the Eulerian solution presented here, but conservatively we estimate that $<\Delta v>$ is within 5\% of the plane parallel shock approximation.


\end{appendix}

\end{document}